\documentclass[prd,preprint,showpacs,groupedaddress]{revtex4-1}
\usepackage{amsmath}
\pdfoutput=1
\usepackage{amsfonts}
\usepackage{amssymb}
\usepackage{float}
\usepackage{geometry}
\usepackage{natbib}
\usepackage[english]{babel}
\usepackage{graphicx}
\usepackage{epstopdf}
\usepackage{subfigure}
\usepackage{caption}
\usepackage{multirow}
\usepackage{indentfirst}
\usepackage{mathrsfs}
\usepackage[colorlinks,linkcolor=blue,anchorcolor=blue,citecolor=blue]{hyperref}

\begin{document}

  \setlength{\parindent}{2em}
  \title{Coexistence of Type-\uppercase\expandafter{\romannumeral2} and Type-\uppercase\expandafter{\romannumeral4} Dirac Fermions in SrAgBi}
  \author{Tian-Chi Ma}
  \author{Jing-Nan Hu}
  \author{Yuan Chen}
  \author{Lei Shao}
  \author{Xian-Ru Hu}
  \author{Jian-Bo Deng} \email[Jian-Bo Deng: ]{dengjb@lzu.edu.cn}
  \affiliation{Department of Physics, Lanzhou University, Lanzhou 730000, China}
  \date{\today}

  \begin{abstract}

  Relativistic massless Weyl and Dirac fermions have isotropic and linear dispersion relations to maintain Poincar\'{e} symmetry, which is the most basic symmetry in high-energy physics. The situation in condensed matter physics is less constrained; only certain subgroups of Poincar\'{e} symmetry --  the 230 space groups that exist in 3D lattices -- need be respected. Then, the free fermionic excitations that have no high energy analogues could exist in solid state systems. Here, We discovered a type of nonlinear Dirac fermion without high-energy analogue in SrAgBi and named it type-\uppercase\expandafter{\romannumeral4} Dirac fermion. The type-\uppercase\expandafter{\romannumeral4} Dirac fermion has a nonlinear dispersion relationship and is similar to the type-\uppercase\expandafter{\romannumeral2} Dirac fermion, which has electron pocket and hole pocket. The effective model for the type-\uppercase\expandafter{\romannumeral4} Dirac fermion is also found. It is worth pointing out that there is a type-\uppercase\expandafter{\romannumeral2} Dirac fermion near this new Dirac fermion.  So we used two models to describe the coexistence of these two Dirac fermions. Topological surface states of these two Dirac points are also calculated. We envision that our findings will stimulate researchers to study novel physics of type-\uppercase\expandafter{\romannumeral4} Dirac fermions, as well as the interplay of type-\uppercase\expandafter{\romannumeral2} and type-\uppercase\expandafter{\romannumeral4} Dirac fermions.

  \end{abstract}

  \maketitle

\section{Introduction}

  Dirac semimetals(DSMs) are currently drawing intense interest in condensed matter and materials physics~\cite{wang2012dirac,liu2014discovery,liu2014stable,liang2015ultrahigh,borisenko2014experimental,chang2017type,huang2017black,milicevic2018tilted,young2012dirac,gibson2015three,young2015dirac,chen2015optical,wieder2016double,novak2015large,zhu2019composite}. They possess four-fold-degenerate Dirac points(DPs) close to the Fermi level.  Dirac point can be regarded as two Weyl points~\cite{wan2011topological,burkov2011weyl,xu2015discovery,weng2015weyl,xu2015discovery,lu2015experimental,lv2015experimental,soluyanov2015type} with opposite chirality meeting at the same $k$-point. The DSMs provide a fertile ground for exploring relativistic particles and high-energy phenomenology at the far more accessible solid-state physics scale. Existing research shows that the Dirac fermions in crystals include linear Dirac fermions and higher-order Dirac fermions.
  
\begin{figure}[h]
  	\centering
  	\subfigure[]{
  		\begin{minipage}[t]{0.3\linewidth}
  			\centering
  			\includegraphics[width=2.5cm]{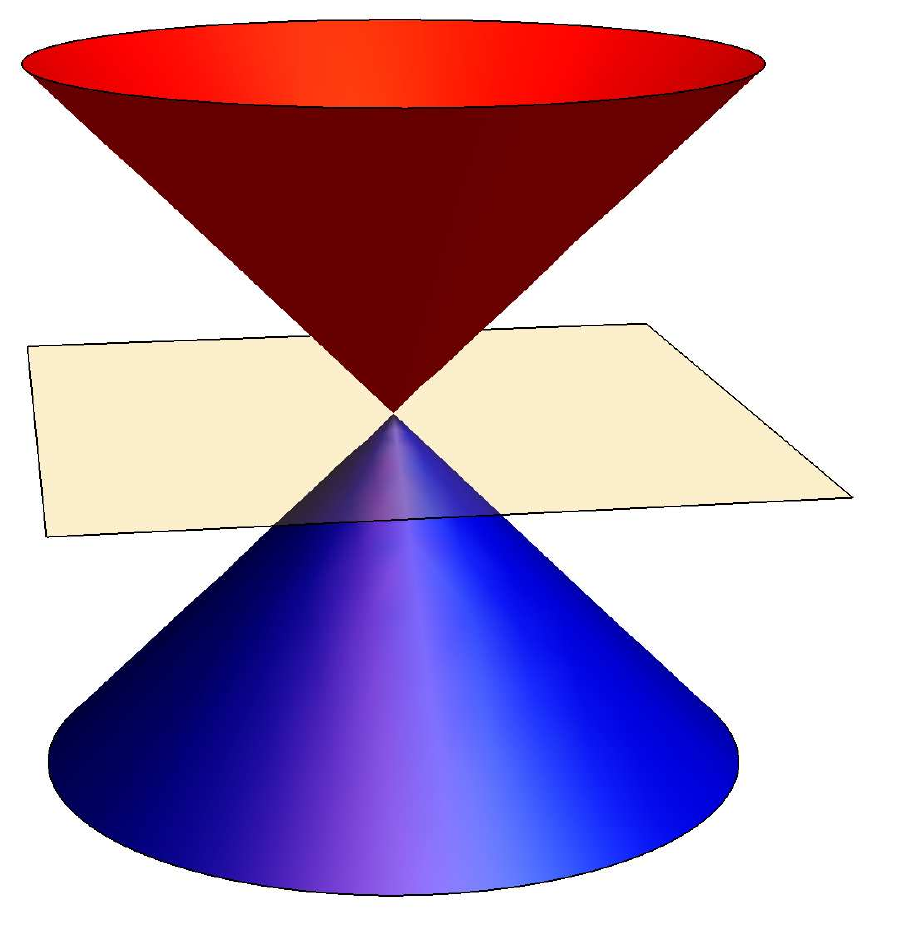}
  		\end{minipage}
  	}
  	\subfigure[]{
  		\begin{minipage}[t]{0.3\linewidth}
  			\centering
  			\includegraphics[width=2.5cm]{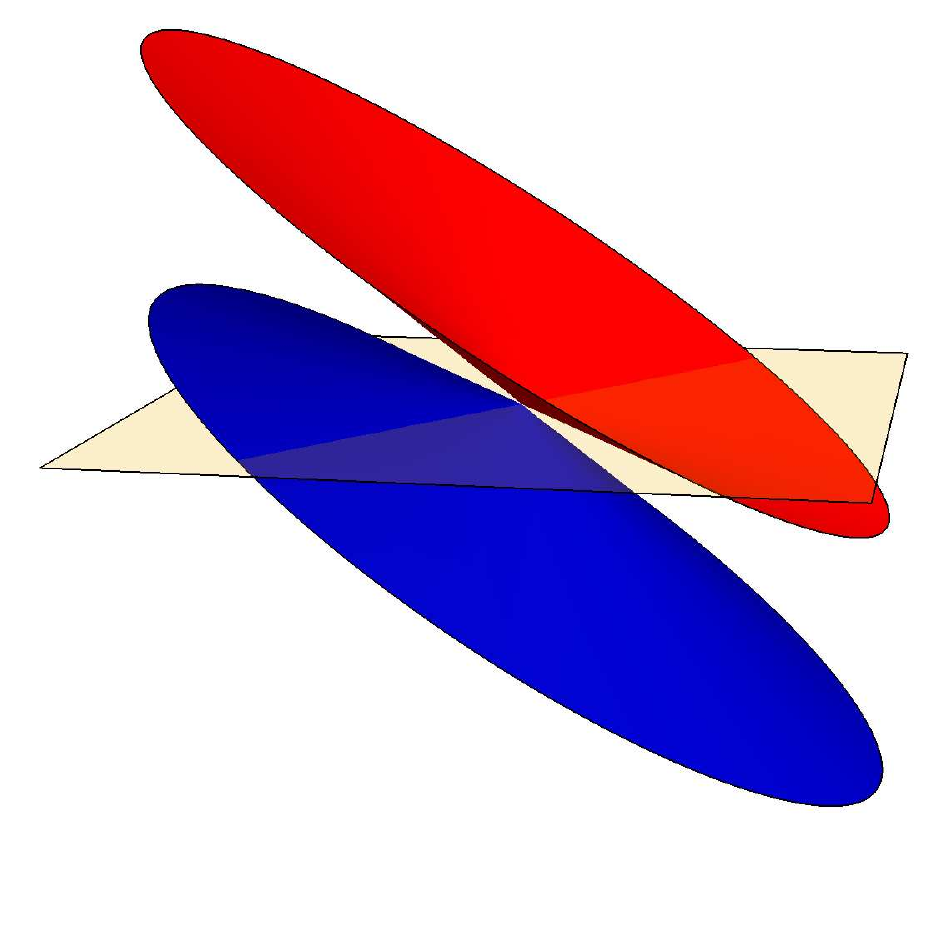}
  		\end{minipage}
  	}
  	\subfigure[]{
  		\begin{minipage}[t]{0.3\linewidth}
  			\centering
  			\includegraphics[width=2.5cm]{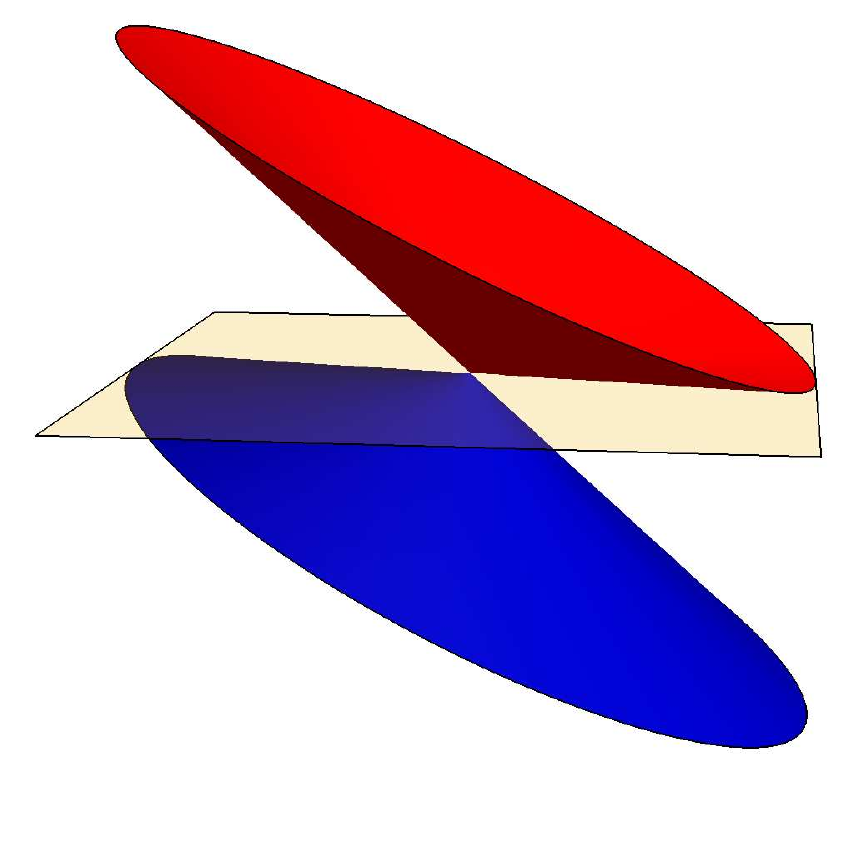}
  		\end{minipage}
  	}
  	\subfigure[]{
  		\begin{minipage}[t]{0.3\linewidth}
  			\centering
  			\includegraphics[width=2.5cm]{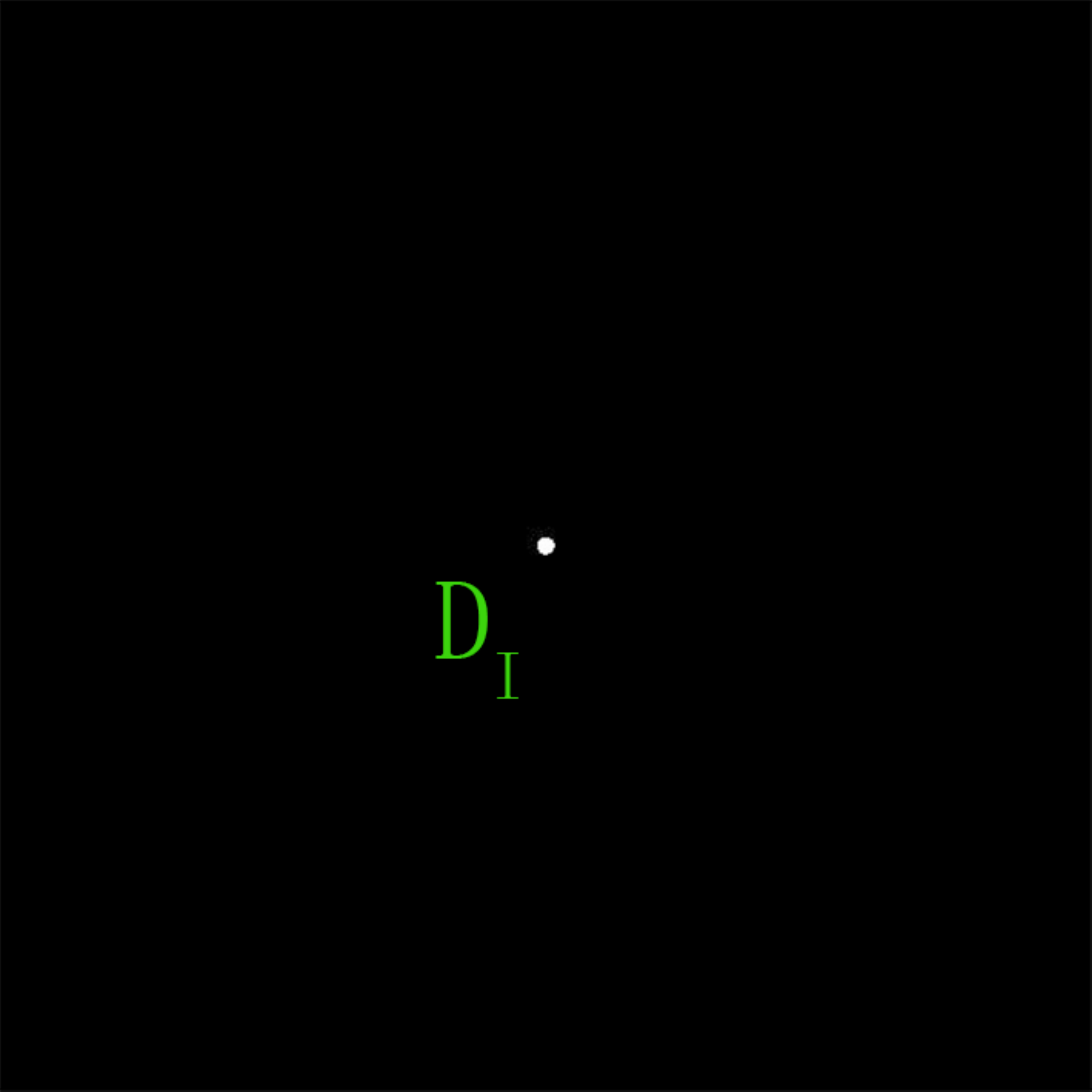}
  		\end{minipage}
  	}
  	\subfigure[]{
  		\begin{minipage}[t]{0.3\linewidth}
  			\centering
  			\includegraphics[width=2.5cm]{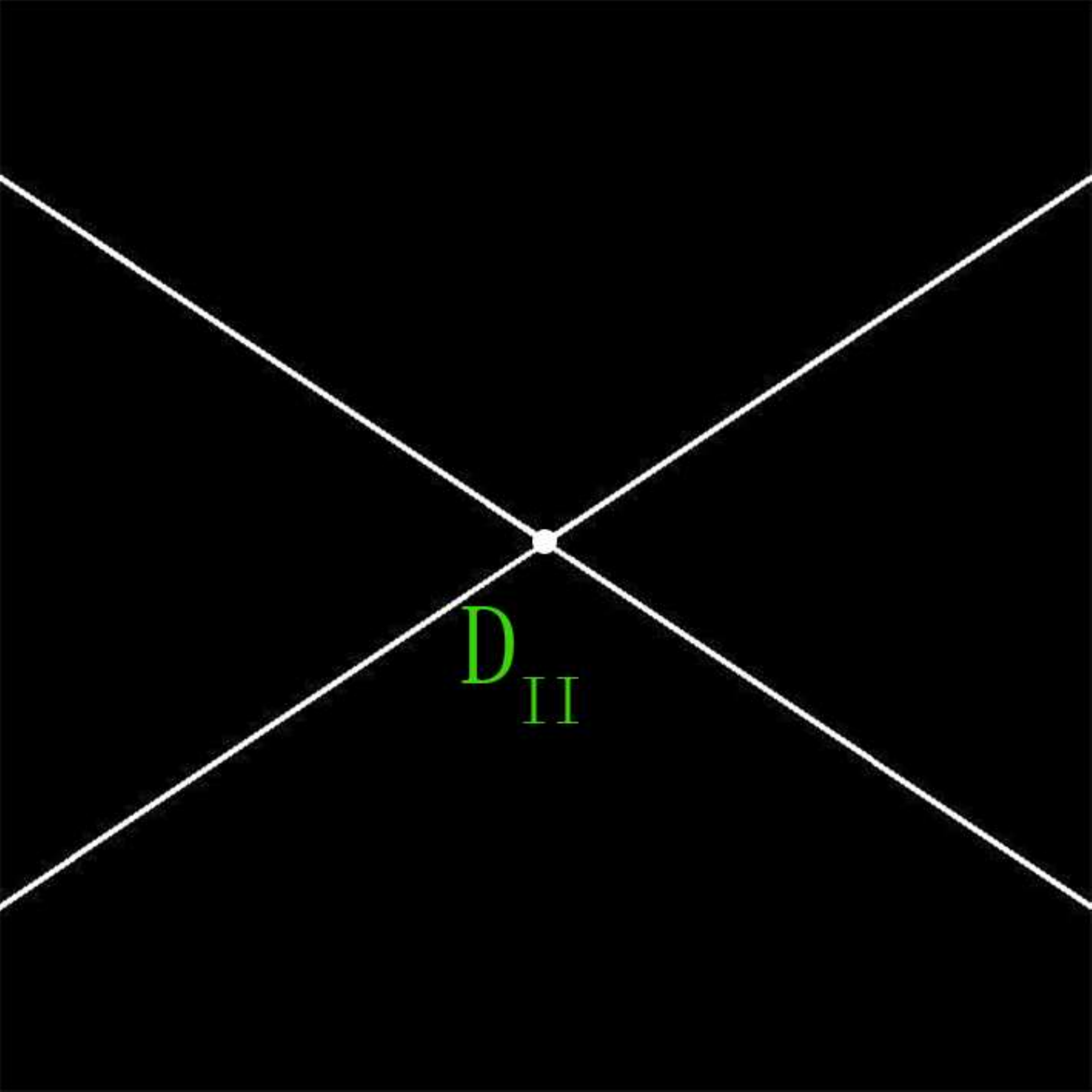}
  		\end{minipage}
  	}
  	\subfigure[]{
  		\begin{minipage}[t]{0.3\linewidth}
  			\centering
  			\includegraphics[width=2.5cm]{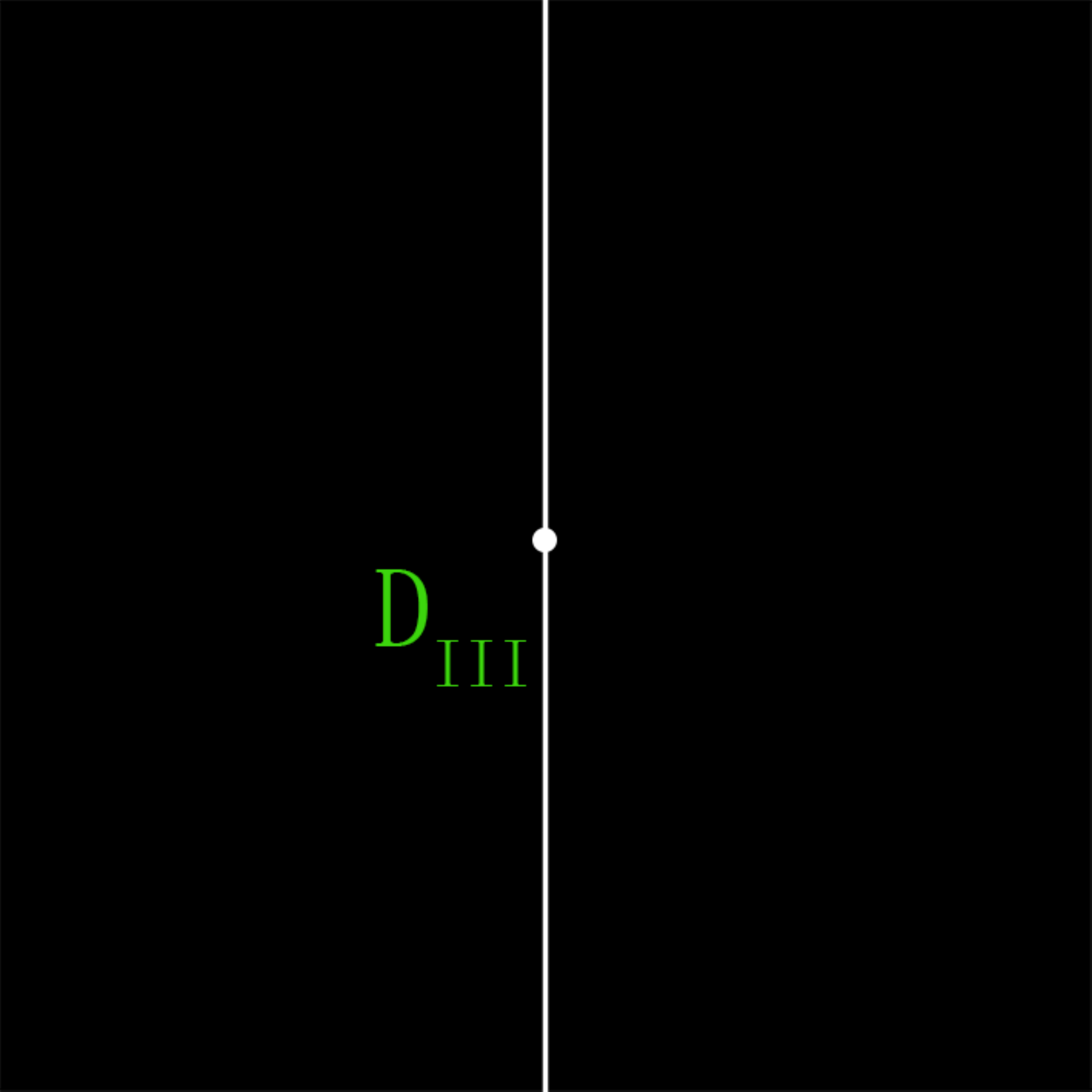}
  		\end{minipage}
  	}
  	\centering
  	\caption{(a) Type-\uppercase\expandafter{\romannumeral1} DP with a point-like Fermi surface. (b) Type-\uppercase\expandafter{\romannumeral2} DP is the contact point between electron and hole pocket. (c) Type-\uppercase\expandafter{\romannumeral3} DP appears as the touching point between Dirac lines. The light yellow semitransparent plane corresponds to the position of the Fermi level. (d)-(f) Fermi surfaces of three types of DSMs.}
  	\label{type 1 to 3}
  \end{figure}

  At present, scientists have discovered three kinds of linear Dirac fermions in crystals. These three kinds of linear Dirac fermions can be described by one model~\cite{huang2017black}:
  \begin{equation}\label{eq:model type 1 to 3}
  	H(\mathbf{k}) =
  	\left[
  	\begin{array}{cc}
  		h(\mathbf{k}) & \mathbf{0} \\
  		\mathbf{0} & h^*(\mathbf{-k})
  	\end{array}
  	\right ]
  \end{equation}
  with
  \begin{equation}
  	h(\mathbf{k})=\mathbf{v}\cdot \mathbf{k}\sigma_{0}+\sum_{i,j}k_{i}A_{ij}\sigma_{j},
  \end{equation}
  where $\sigma_{0}$ is the identity matrix and $\sigma_{j}$ are Pauli matrices. The energy spectrum of a DP is
  \begin{equation}
  	\begin{aligned}
  		E_{\pm}(\mathbf{k}) & = \sideset{}{_i}\sum v_{i}k_{i} \pm \sqrt{\sideset{}{_j}\sum (\sideset{}{_i}\sum k_{i}A_{ij})^2} \\
  		& = T(\mathbf{k}) \pm U(\mathbf{k}).
  	\end{aligned}
  \end{equation}
  It is known that the band crossing point is a type-\uppercase\expandafter{\romannumeral2} DP if there exist a direction for which $T > U$~\cite{chang2017type,yan2017lorentz,huang2016type,zhang2017experimental}, otherwise it is a type-\uppercase\expandafter{\romannumeral1} DP~\cite{wang2012dirac,liu2014stable,xiao2017manipulation,liu2019engineering,li2014gapless,ulstrup2014ultrafast}. If and only if for a particular direction $\hat{k}$ in reciprocal space, $T(\hat{k}) = U(\hat{k})$, but $T(\hat{k}) < U(\hat{k})$ for other directions, the DPs are connected by a line-like Fermi surface which is the Dirac line of the type-\uppercase\expandafter{\romannumeral3} DSM~\cite{huang2017black,milicevic2019type,jin2020hybrid,gong2020theoretical}. The bands at the energy of DPs and the corresponding Fermi surfaces are shown in Fig. \ref{type 1 to 3}. One can find that the bands of these three types of Dirac semimetals near the DPs are all cones and the Fermi surface are Linear(For type-\uppercase\expandafter{\romannumeral1} DSM, Fermi surface degenerates to a point).

  The Dirac points with nonlinear (higher-order) energy dispersion have also been studied in several works. Zihao Gao $et\,al$. studied classification of stable Dirac and Weyl semimetals with reflection and rotational symmetry, they pointed out that there are two kinds of Dirac semimetals created via ABC and TBC~\cite{gao2016classification}. Bohm-Jung Yang and Naoto Nagaosa propose a framework to classify three dimensional (3D) DSMs in systems having the time-reversal, inversion and uniaxial rotational symmetries, they found that quadratic Dirac point and cubic Dirac point can exist in systems having $C_6$ rotational symmetry with respect to the $z$ axis~\cite{yang2014classification}. Weikang Wu $et\,al$. perform a systematic search over all 230 space groups with time-reversal symmetry and spin-orbit coupling considered, they found that the order of dispersion cannot be higher than three, i.e., only the quadratic and cubic Dirac points (QDPs and CDPs) are possible~\cite{wu2020higher}. Wing Chi Yu $et\,al$. show that a cubic Dirac point involving a threefold axis can be realized close to the Fermi level in the non-ferroelectric phase of LiOs$\mathrm{O}_3$~\cite{yu2018nonsymmorphic}. Qihang Liu and Alex Zunger find that only materials in two space groups, P$6_3/$m (No. 176) and P$6/$mcc (No. 192), have the potential to host cubic Dirac fermions~\cite{liu2017predicted}.
  
  Recently, scientists discovered type-\uppercase\expandafter{\romannumeral3} Weyl points in (TaSe$_4$)$_2$I~\cite{li2019type}. In (TaSe$_4$)$_2$I, the Fermi surface near the Weyl point consists of two electron or two hole pockets and these two pockets are from the same band and the band looks like a saddle. This discovery inspires us whether a similar situation exists in DSMs.

  In this work, we investigate type-\uppercase\expandafter{\romannumeral2} and type-\uppercase\expandafter{\romannumeral4} Dirac fermions coexist in SrAgBi. Scientists' research interest in SrAgBi family materials has never been reduced~\cite{gibson2015three,chen2017ternary,chen2017hybrid,sasmal2020magnetotransport,chaiconfirmation,mardanya2019prediction,xu2020crystal,nakayama2020observation,tai2020anisotropic}. SrAgBi family materials were first predicted to exist DPs in Ref. \cite{gibson2015three}. Later, CaAgBi was predicted to coexist type-\uppercase\expandafter{\romannumeral1} and type-\uppercase\expandafter{\romannumeral2} Dirac fermions in Ref. \cite{chen2017ternary}. Here, we studied a new type of Dirac fermions in SrAgBi and is dubbed type-\uppercase\expandafter{\romannumeral4} Dirac fermions. At the energy of type-\uppercase\expandafter{\romannumeral4} Dirac fermions, the Fermi surface consists of a electron pocket and a hole pocket, it is similar to type-\uppercase\expandafter{\romannumeral2} Dirac Fermions, but the bands are non-linear. The electron pocket is in the hole pocket and they touch at the DPs.  The Fermi arcs of both DPs were clearly discovered. Meanwhile, we use a eight-band $k{\cdot}p$ model to discribe the bandcrossing near the Fermi level along $\Gamma$-A. More importantly, we reproduce the  bands of SrAgBi near the Fermi level with a tight-binding model. The Fermi surface of this model is consistent with the result of first-principles band-structure calculations.

\section{COMPUTATION METHODS}
In this paper, density-functional theory (DFT) calculations with the projected augmented wave (PAW) are implemented in the VASP~\cite{perdew1981self,kresse1996efficiency} with generalized gradient approximation (GGA)~\cite{perdew1996generalized}. The spin-orbit coupling (SOC) was employed in the electronic structure calculations. The tight-binding model matrix elements are calculated by projecting onto the Wannier orbitals~\cite{marzari1997maximally,souza2001maximally,mostofi2014updated}. We used Sr $d$-orbitals, Ag $d$-orbitals and Bi $p$-orbitals as initial wave functions for maximizing localization. Surface spectra were calculated based on the iterative Green¡¯s function with the help of WannierTools~\cite{wu2018wanniertools}. The eight-band tight-binding model is calculated by PythTB~\cite{yusufaly2013tight}.

\section{RESULTS}
\subsection{Electronic structure}
\begin{figure}[h]
	\centering
	\subfigure[]{
		\begin{minipage}[t]{0.46\linewidth}
			\centering
			\includegraphics[width=3.5cm]{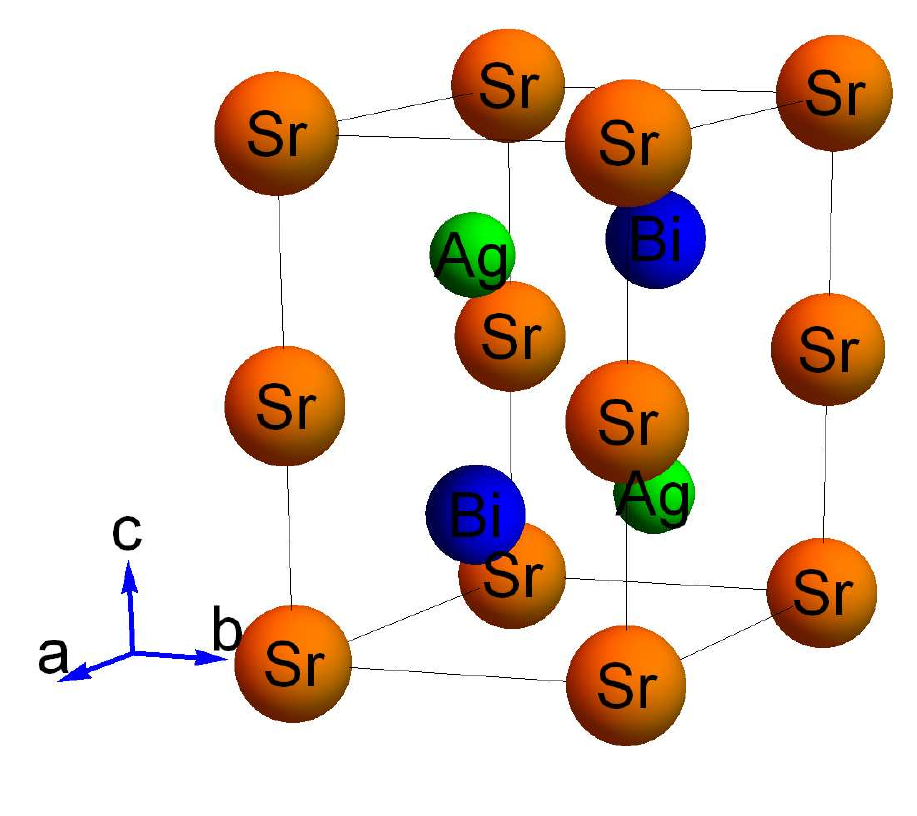}
		\end{minipage}
	}
	\subfigure[]{
		\begin{minipage}[t]{0.46\linewidth}
			\centering
			\includegraphics[width=3.5cm]{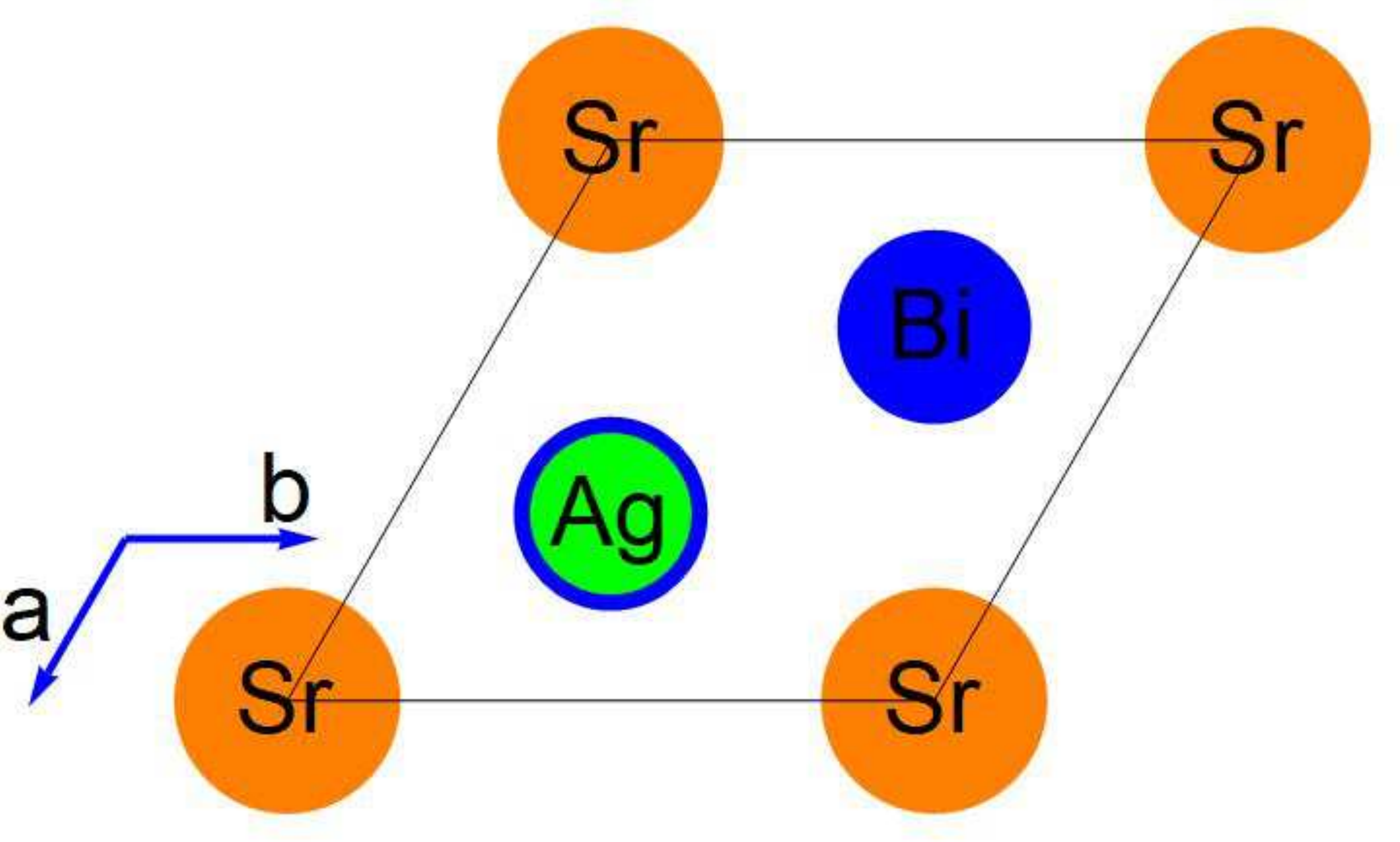}
		\end{minipage}
	}
	\centering
	\caption{(a) Side view and (b) top view of the crystal structure of SrAgBi.}
	\label{bravais lattice}
\end{figure}

\begin{figure}[htp]
	\centering
	\subfigure[]{
		\begin{minipage}[t]{0.9\linewidth}
			\centering
			\includegraphics[width=7.0cm]{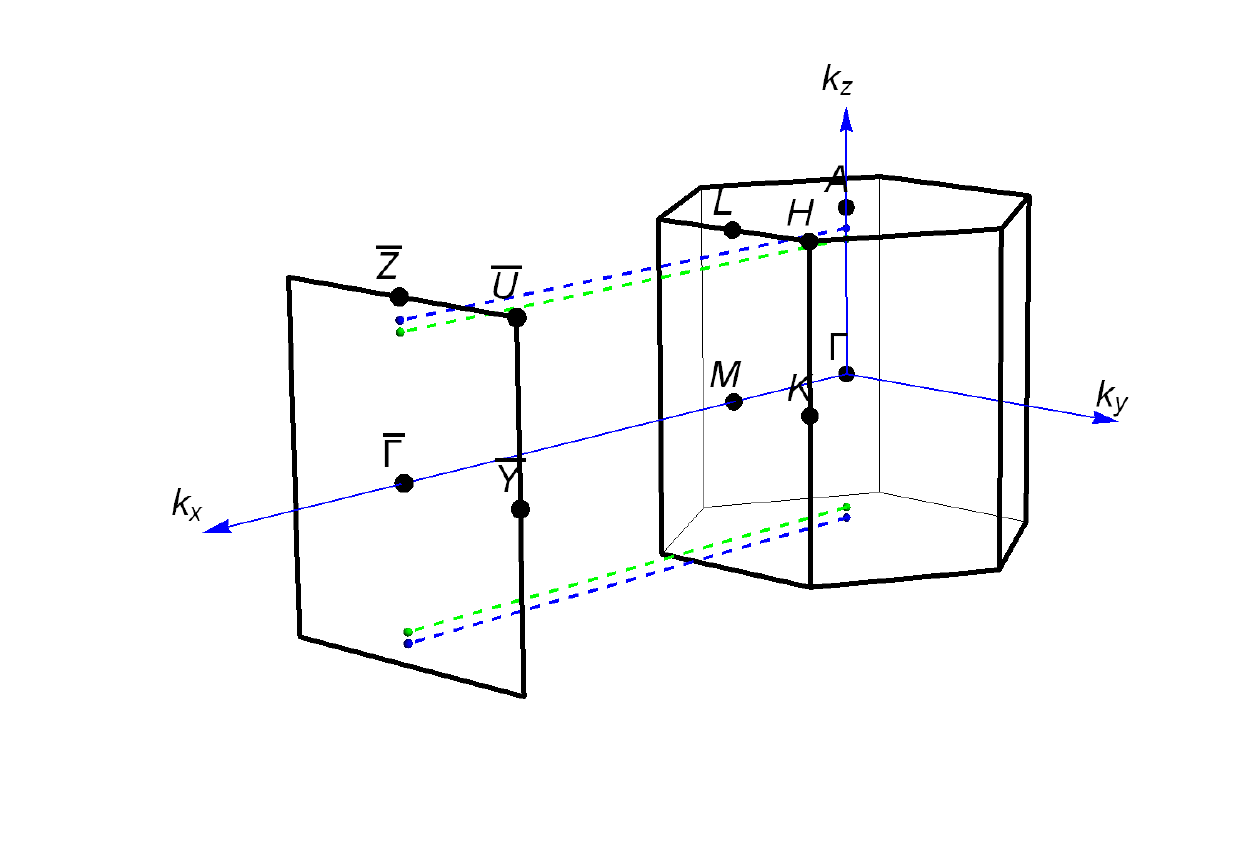}
		\end{minipage}
	}
	\subfigure[]{
		\begin{minipage}[t]{0.9\linewidth}
			\centering
			\includegraphics[width=5.0cm]{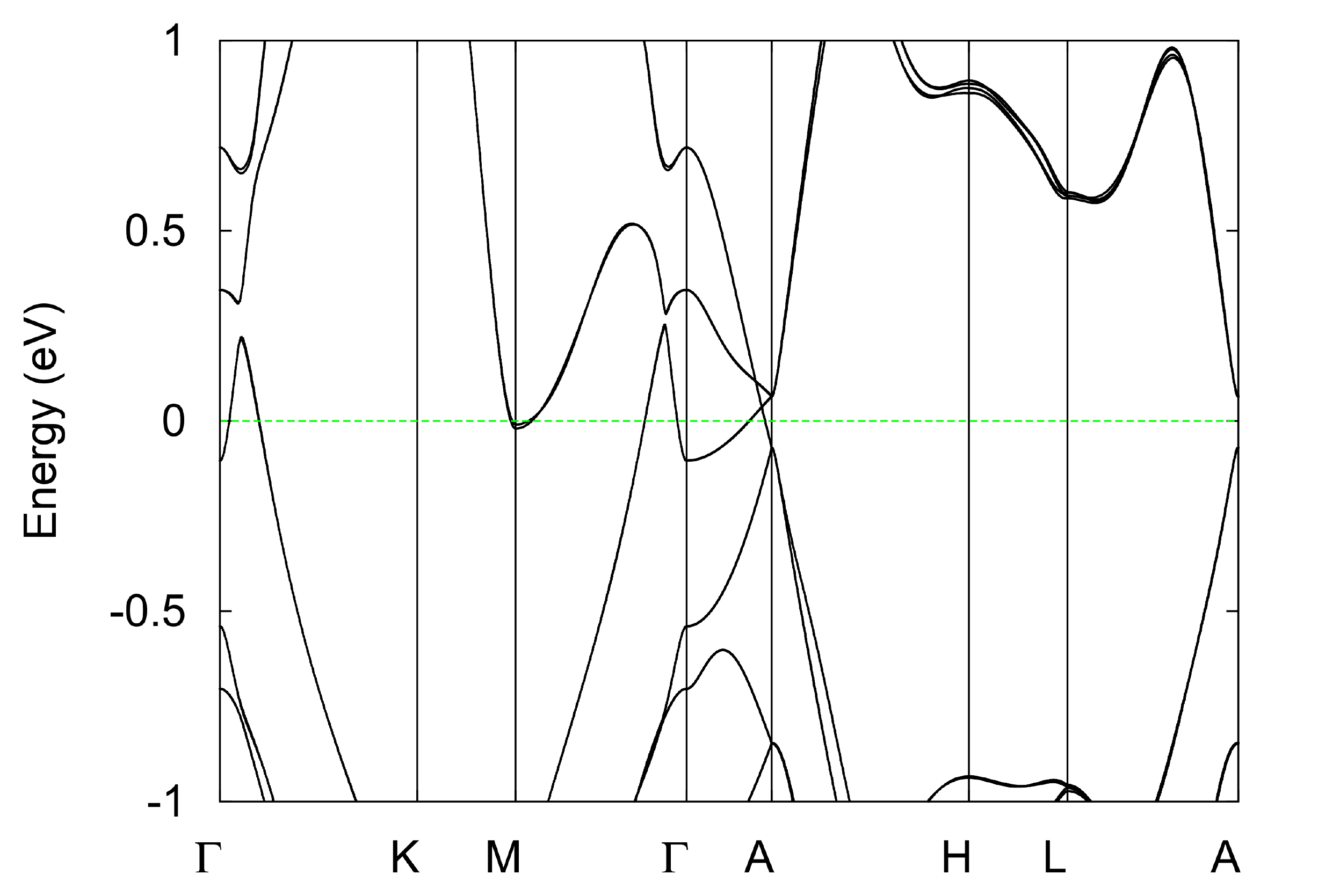}
		\end{minipage}
	}
	\subfigure[]{
		\begin{minipage}[t]{0.46\linewidth}
			\centering
			\includegraphics[width=3cm,height=2cm]{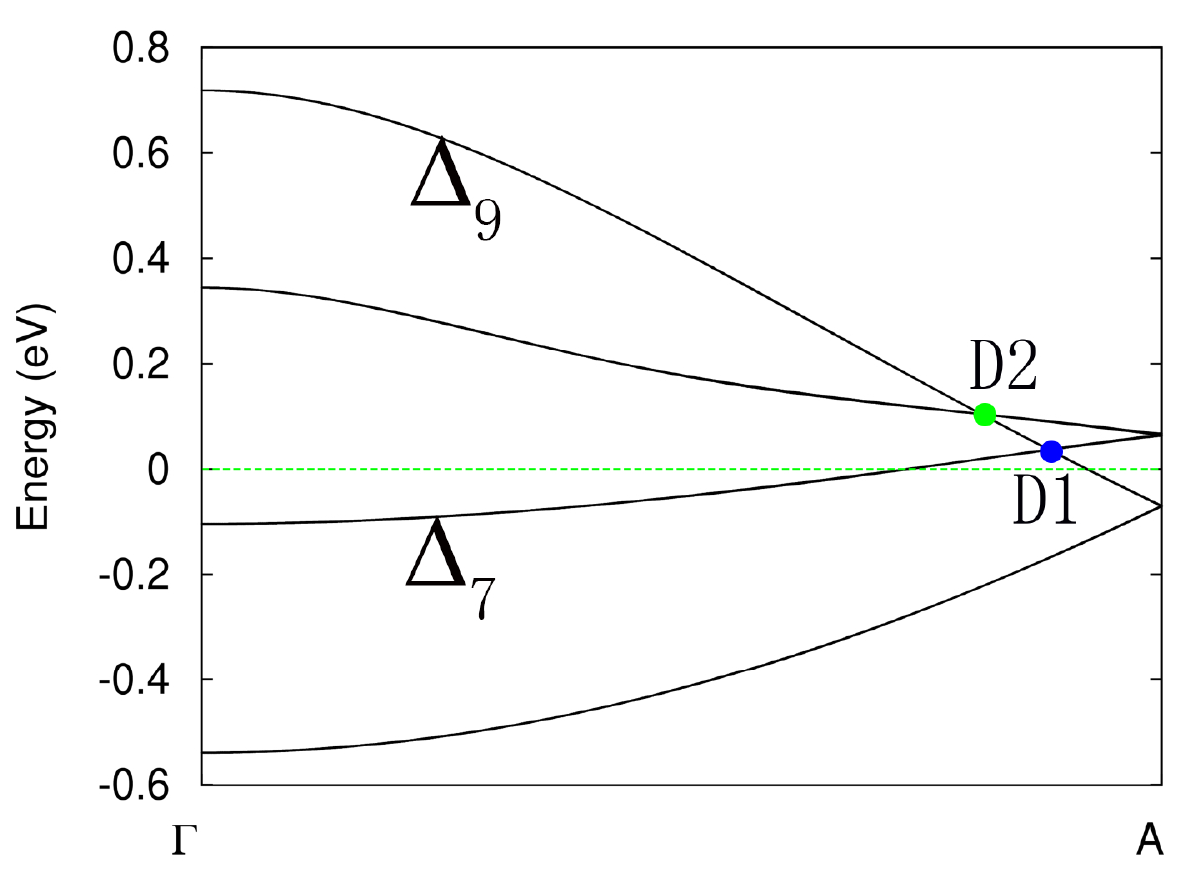}
		\end{minipage}
	}
	\subfigure[]{
		\begin{minipage}[t]{0.46\linewidth}
			\centering
			\includegraphics[width=3cm]{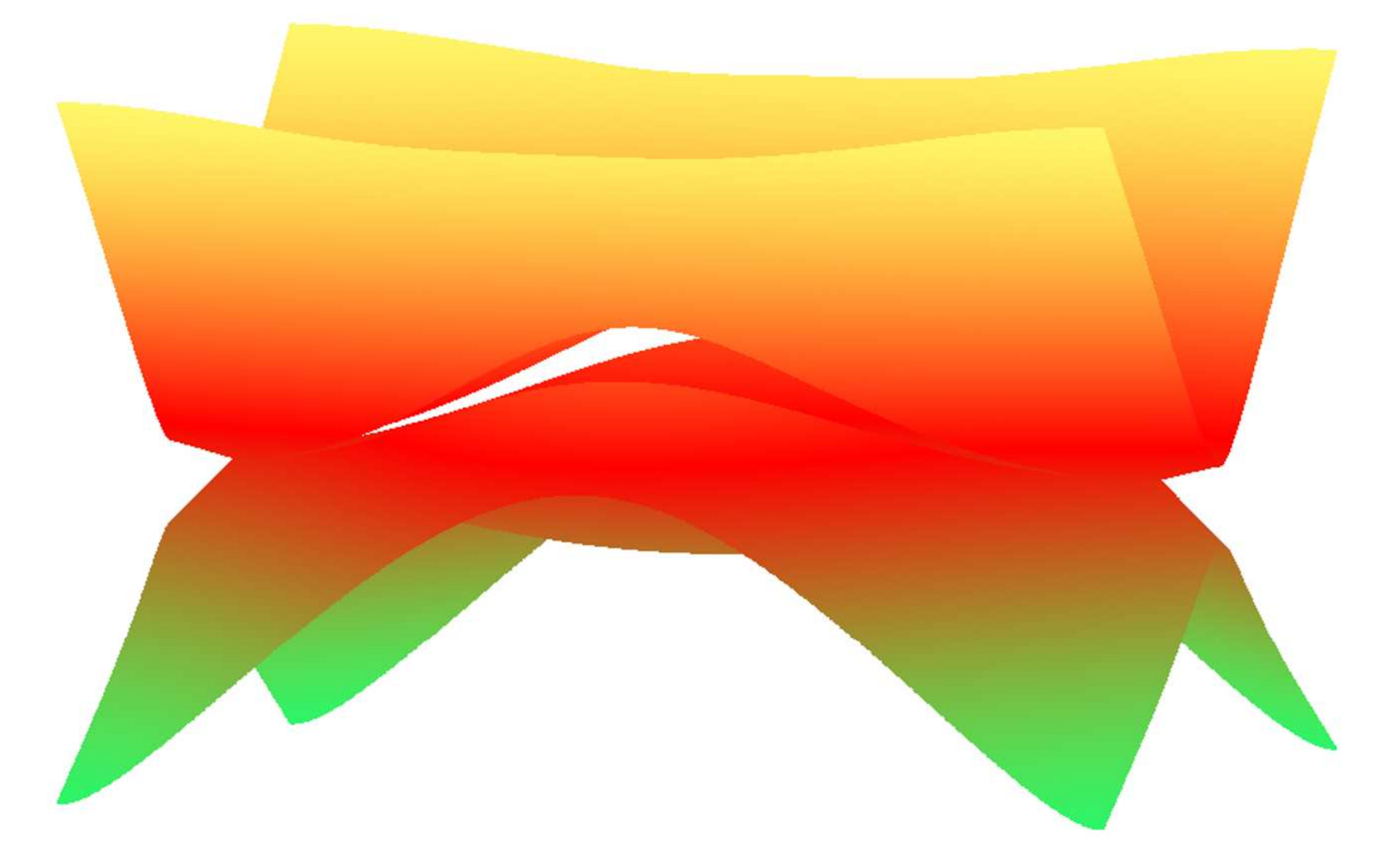}
		\end{minipage}
	}
	\subfigure[]{
		\begin{minipage}[t]{0.46\linewidth}
			\centering
			\includegraphics[width=4cm]{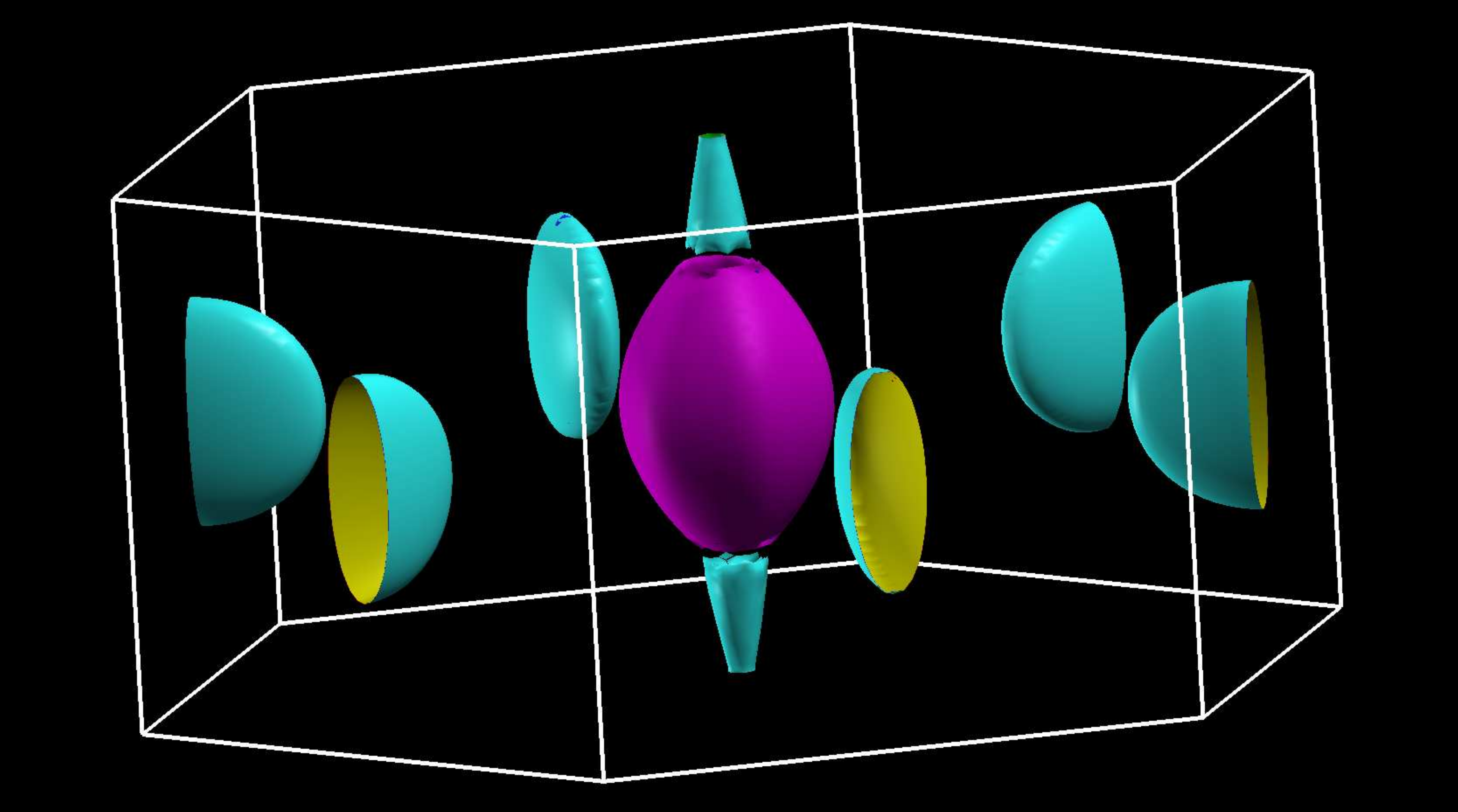}
		\end{minipage}
	}
	\centering
	\caption{(a) BZ of bulk and the projected surface BZ of $(100)$ plane. Band structure on high symmetry line (b), (c) and $k_y-k_z$ plane (d). (e) Bulk-state equal-energy contours of E = 0.1043 eV.}
	\label{bands and pocket}
\end{figure}

The SrAgBi family of compounds crystalizes in a hexagonal Bravais lattice with space group $D_{6h}^{4}$ ($P6_{3}/mmc$, No. 194). As shown in Figs. \ref{bravais lattice}(a) and \ref{bravais lattice}(b). The crystal structure can be viewed as stacked graphene layers; the Sr$^{2+}$ cations are stacked between [Ag$^{1+}$As$^{3-}$]$^{2-}$ in the honeycomb network. Fig. \ref{bands and pocket}(a) shows the bulk and surface Brillouin zone (BZ) of SrAgBi  crystal. The type-\uppercase\expandafter{\romannumeral2} and type-\uppercase\expandafter{\romannumeral4} DPs are marked by green points and blue points, respectively.

\begin{figure*}[htp]
	\centering
	\subfigure[]{
		\begin{minipage}[t]{0.3\linewidth}
			\centering
			\includegraphics[width=6cm,height=5cm]{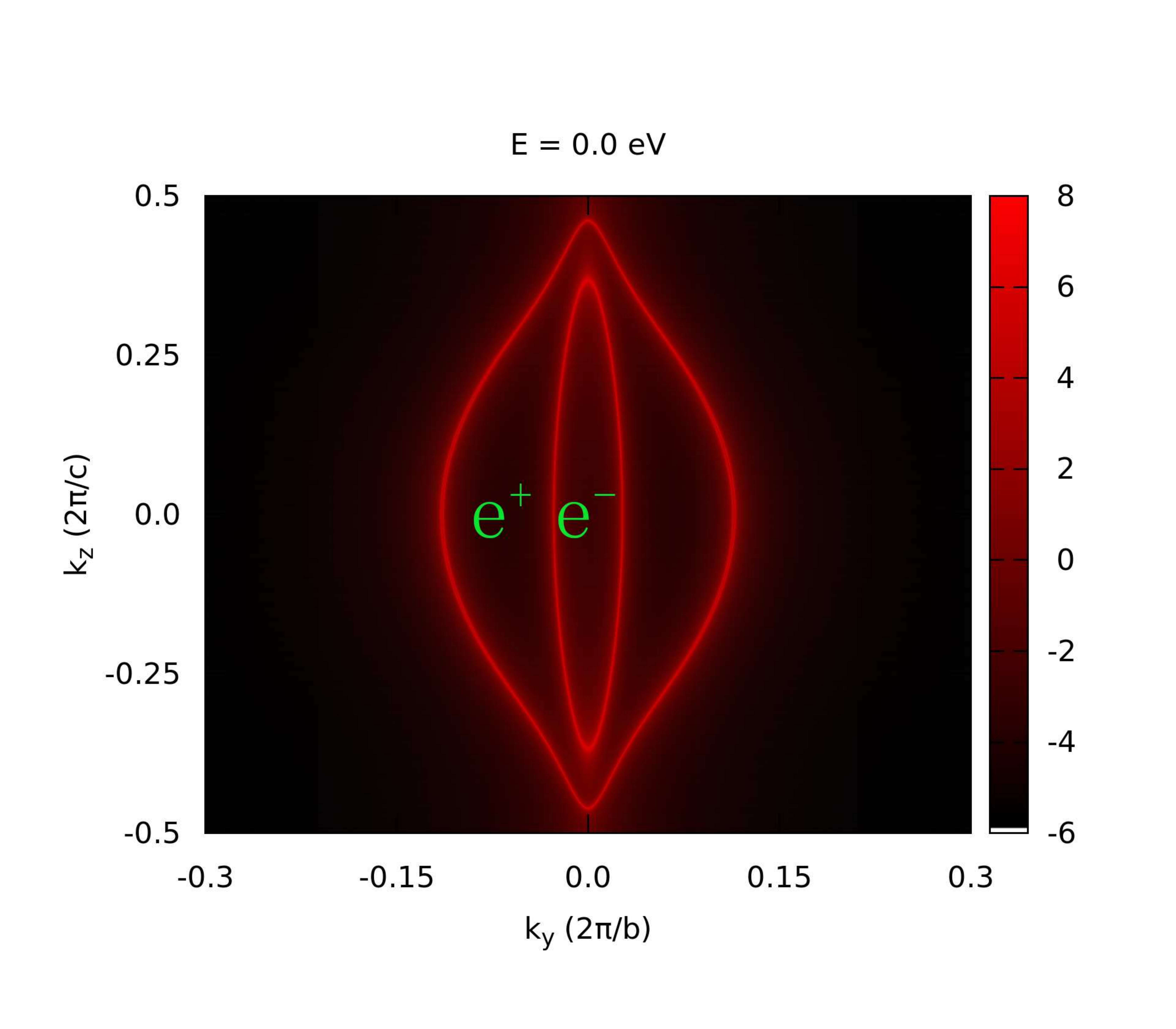}
		\end{minipage}
	}
	\subfigure[]{
		\begin{minipage}[t]{0.3\linewidth}
			\centering
			\includegraphics[width=6cm,height=5cm]{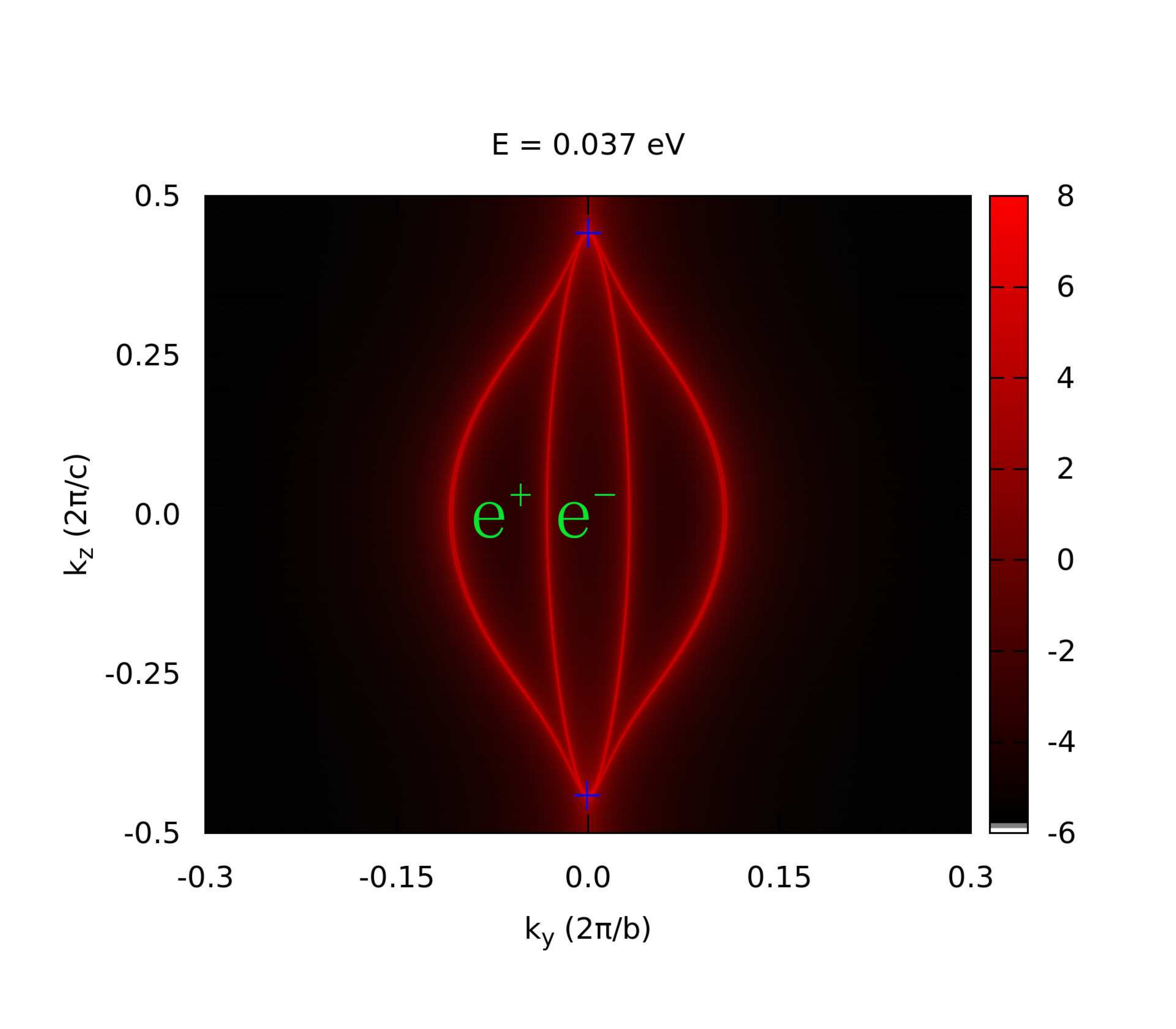}
		\end{minipage}
	}
	\subfigure[]{
		\begin{minipage}[t]{0.3\linewidth}
			\centering
			\includegraphics[width=6cm,height=5cm]{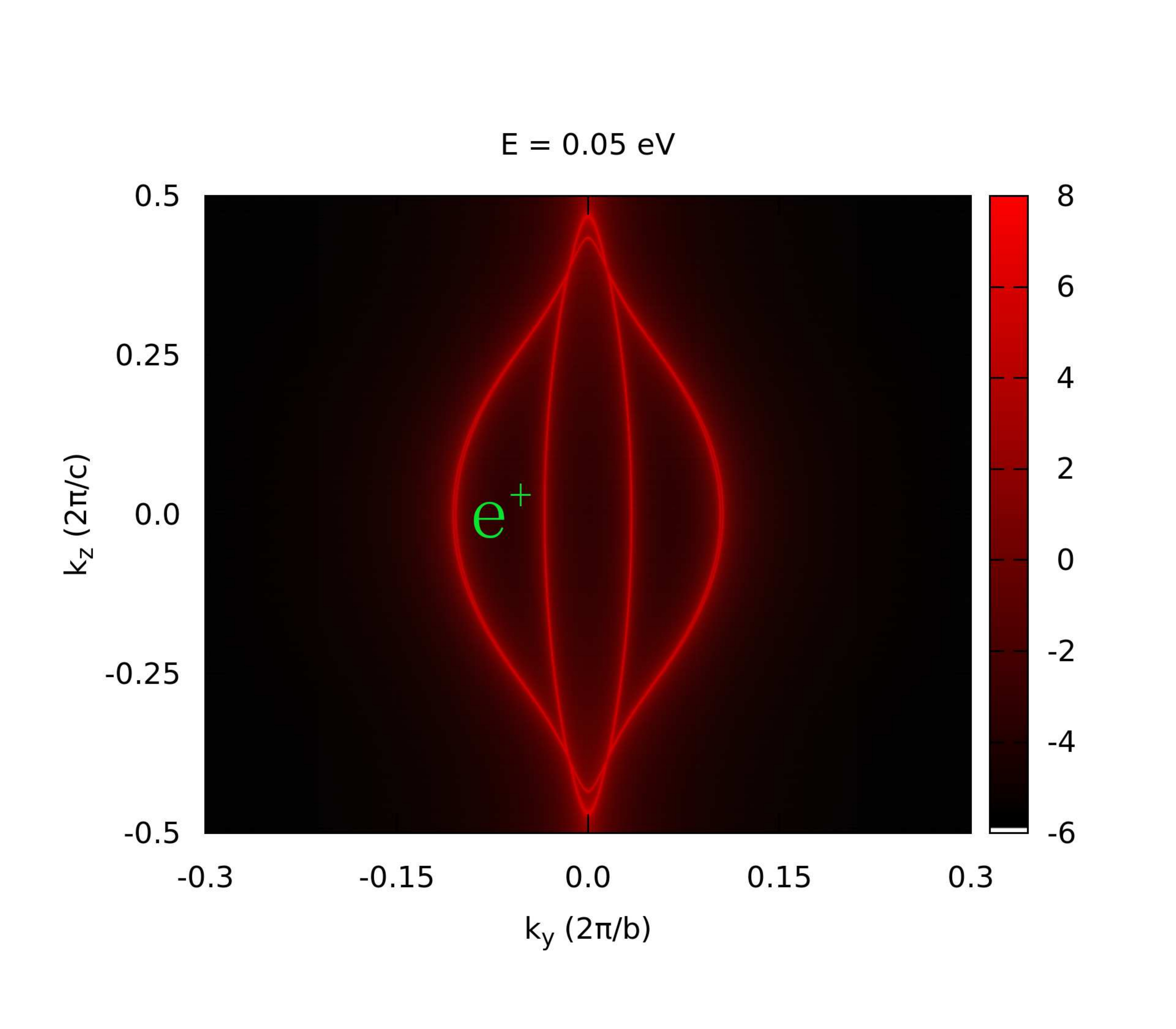}
		\end{minipage}
	}
	\subfigure[]{
		\begin{minipage}[t]{0.3\linewidth}
			\centering
			\includegraphics[width=6cm,height=5cm]{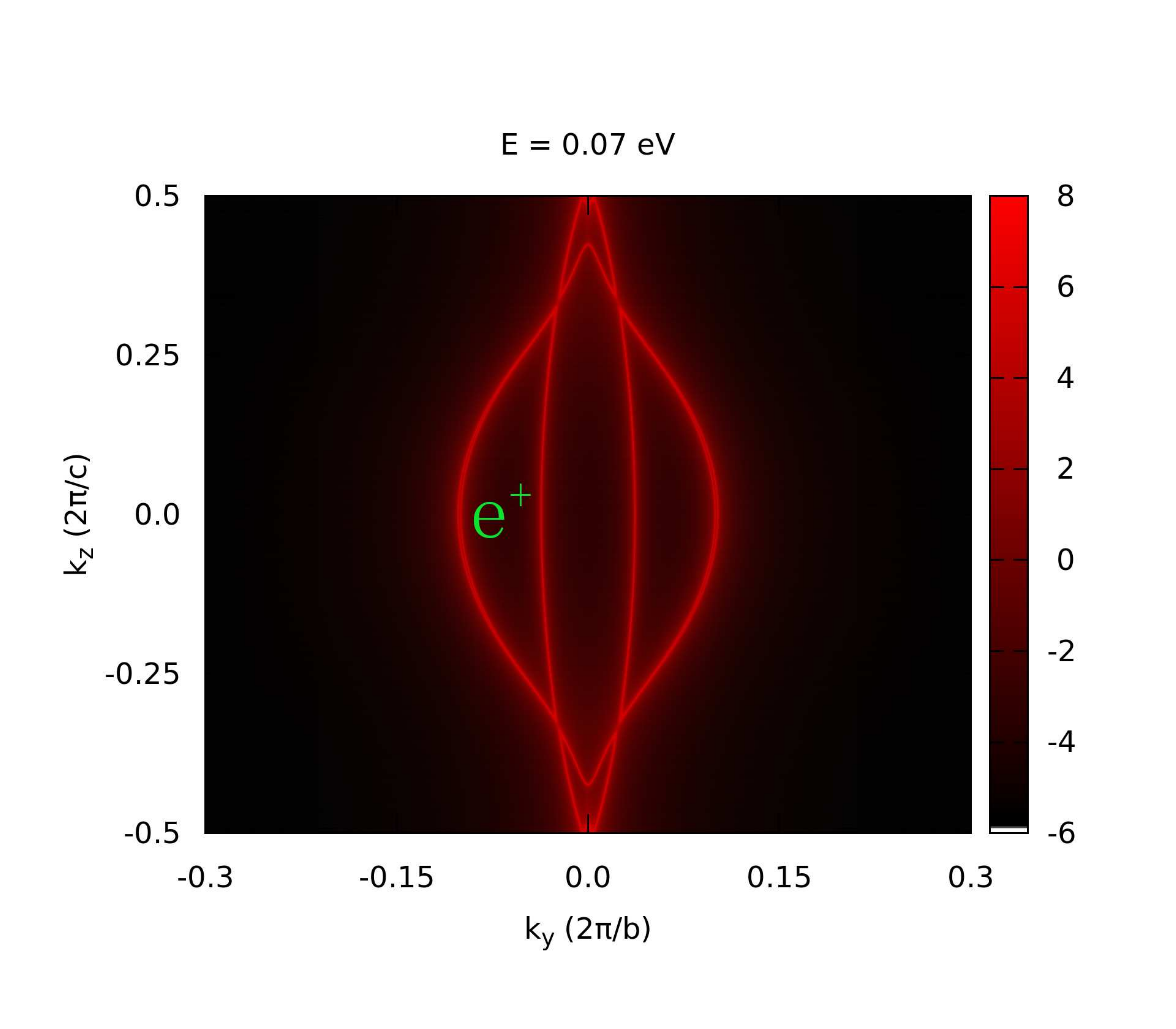}
		\end{minipage}
	}
	\subfigure[]{
		\begin{minipage}[t]{0.3\linewidth}
			\centering
			\includegraphics[width=6cm,height=5cm]{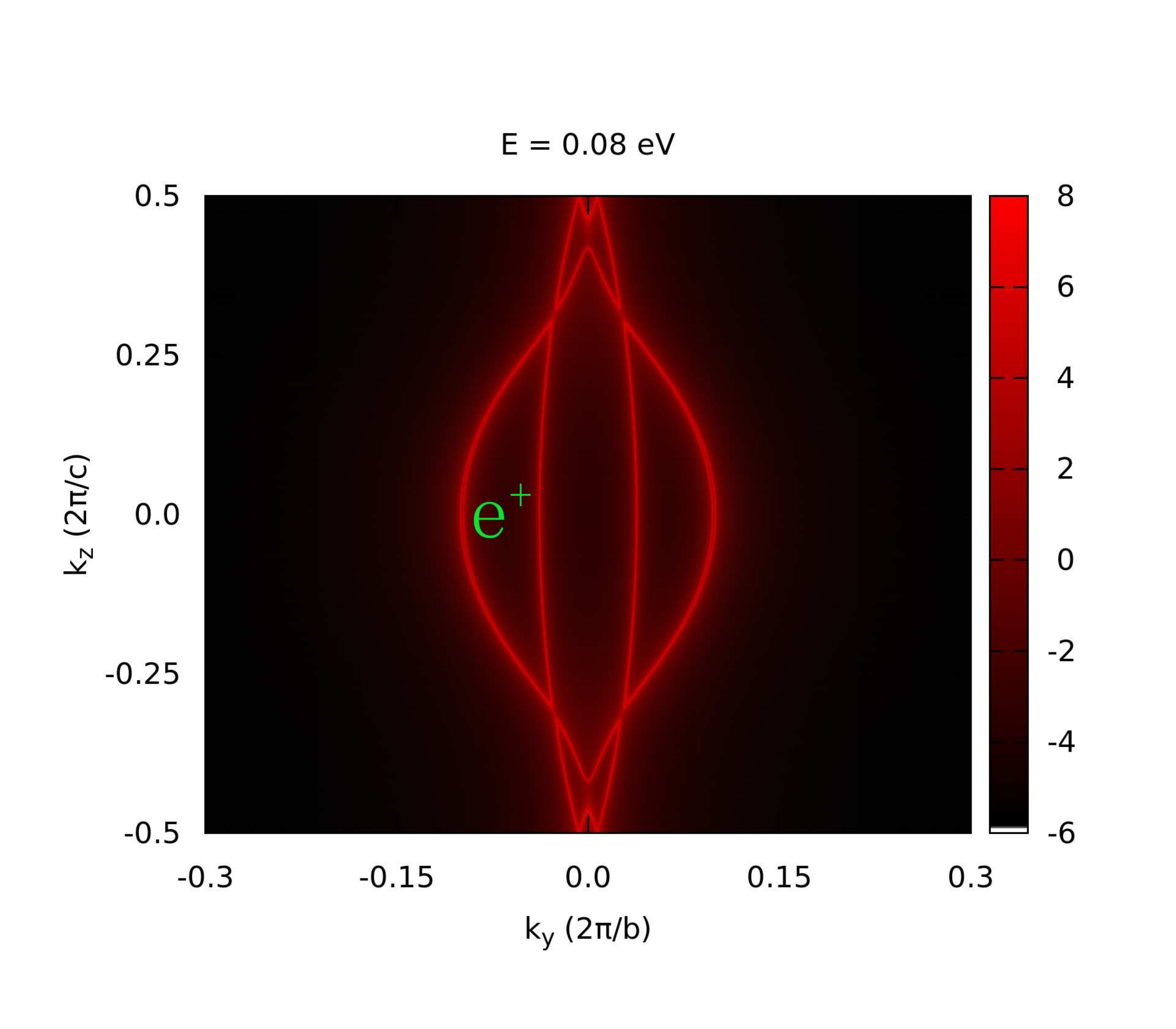}
		\end{minipage}
	}
	\subfigure[]{
		\begin{minipage}[t]{0.3\linewidth}
			\centering
			\includegraphics[width=6cm,height=5cm]{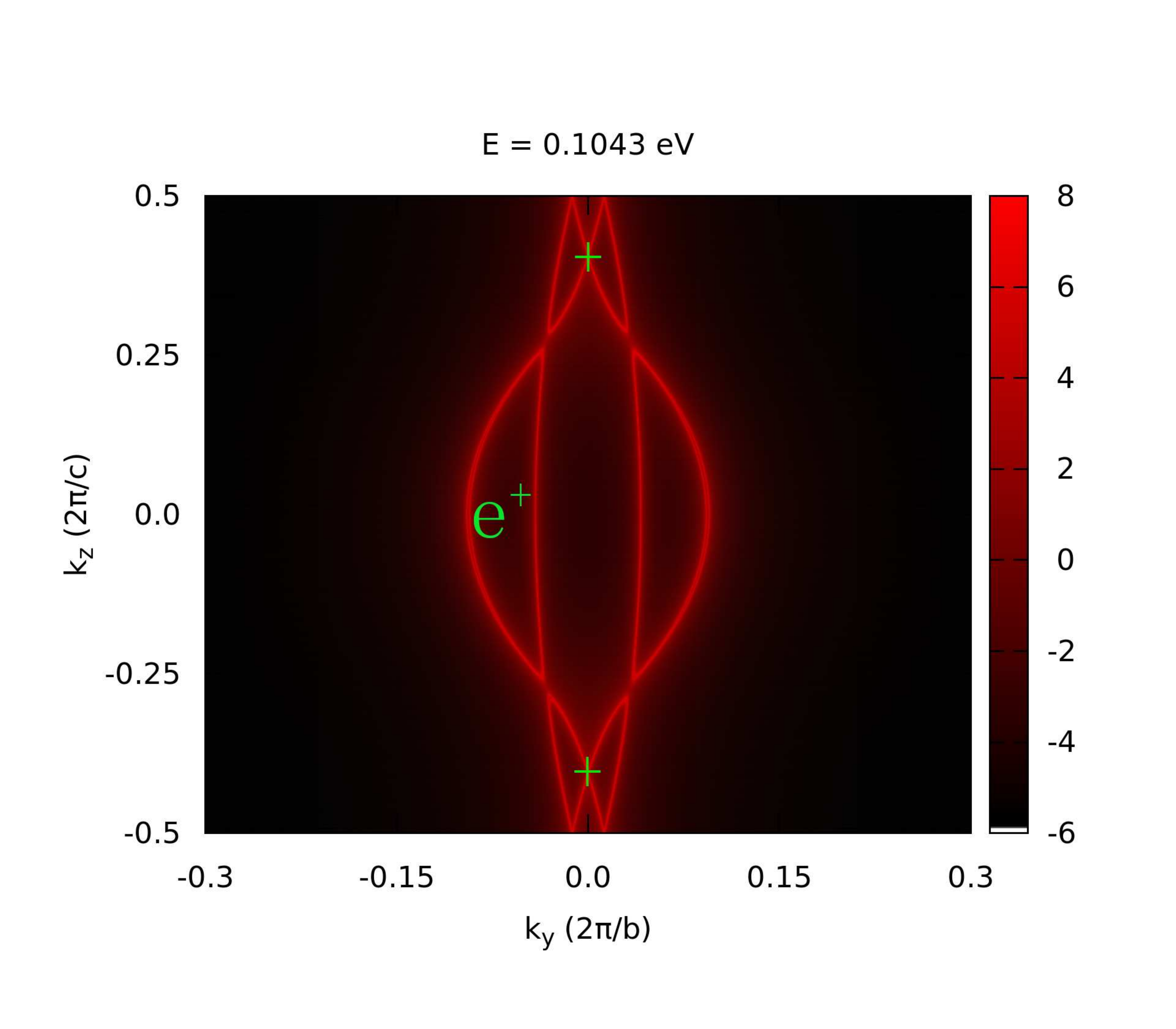}
		\end{minipage}
	}
	\centering
	\caption{Bulk constant energy contours in $(k_y,k_z)$ space at $k_x = 0$ of different energy around the Fermi level.}
	\label{2d pocket 6}
\end{figure*}

We now present the calculated band structure of SrAgBi to reveal the DPs and its type-\uppercase\expandafter{\romannumeral4} character. The calculated bulk band structure along high symmetry directions [Fig. \ref{bands and pocket}(b)] reveals the semimetallic ground state. The band structure shows that there are two DPs along the $\Gamma$-A direction. We mark these two DPs in Fig. \ref{bands and pocket}(c). The blue one is type-\uppercase\expandafter{\romannumeral4} DP and the green one is type-\uppercase\expandafter{\romannumeral2} DP. Figure \ref{bands and pocket}(d) shows the 3D band structure of valance band and conduction band and they touch at type-\uppercase\expandafter{\romannumeral4} DPs. One can find that the band structure near a type-\uppercase\expandafter{\romannumeral4} DP isn't a perfect cone and can't be described by equation (\ref{eq:model type 1 to 3}).  Because a type-\uppercase\expandafter{\romannumeral4} Dirac node exists between $\Gamma$ and A points, and the two planes $k_z$ = 0, $k_z$ = $\pi$ are fully gapped, the two dimensions (2D) topological invariants of the two planes must be topologically distinct. For the illustration, one can consider their 2D Z2 invariants~\cite{kane2005z}, $\nu_0$ and $\nu_\pi$. Using the Fu-Kane method~\cite{fu2007topological}, one results that ($\nu_0$, $\nu_\pi$) = (1, 0). These two invariants indicate the single band inversion at $\Gamma$.

Because the type-\uppercase\expandafter{\romannumeral2} DP and the type-\uppercase\expandafter{\romannumeral4} DP are formed by the intersection of three adjacent bands and they are very close to each other in BZ, We think it is meaningful to study the relationship between these two points. Fig. \ref{2d pocket 6} shows the evolution of bulk-state equal-energy contours for different values of E, the blue ``+" is the position of type-\uppercase\expandafter{\romannumeral4} DP and the green ``+" is the position of type-\uppercase\expandafter{\romannumeral2} DP. $e^+$ and $e^-$ represent hole pocket and electron pocket, respectively. When E = 0.0 eV [Fig. \ref{2d pocket 6}(a)], there is a electron pocket in a hole pocket. When E = 0.037 eV [Fig. \ref{2d pocket 6}(b)], the energy of type-\uppercase\expandafter{\romannumeral4} DP, the electron pocket touches the hole pocket. The touched points are type-\uppercase\expandafter{\romannumeral4} DPs. When the energy continues to increase, the electron pocket disappears and when E = 0.1043 eV [Fig. \ref{2d pocket 6}(f)], the other two pockets touch at type-\uppercase\expandafter{\romannumeral2} DPs.

\subsection{Fermi arcs and surface states}
\begin{figure}[]
	\centering
	\includegraphics[width=.49\textwidth]{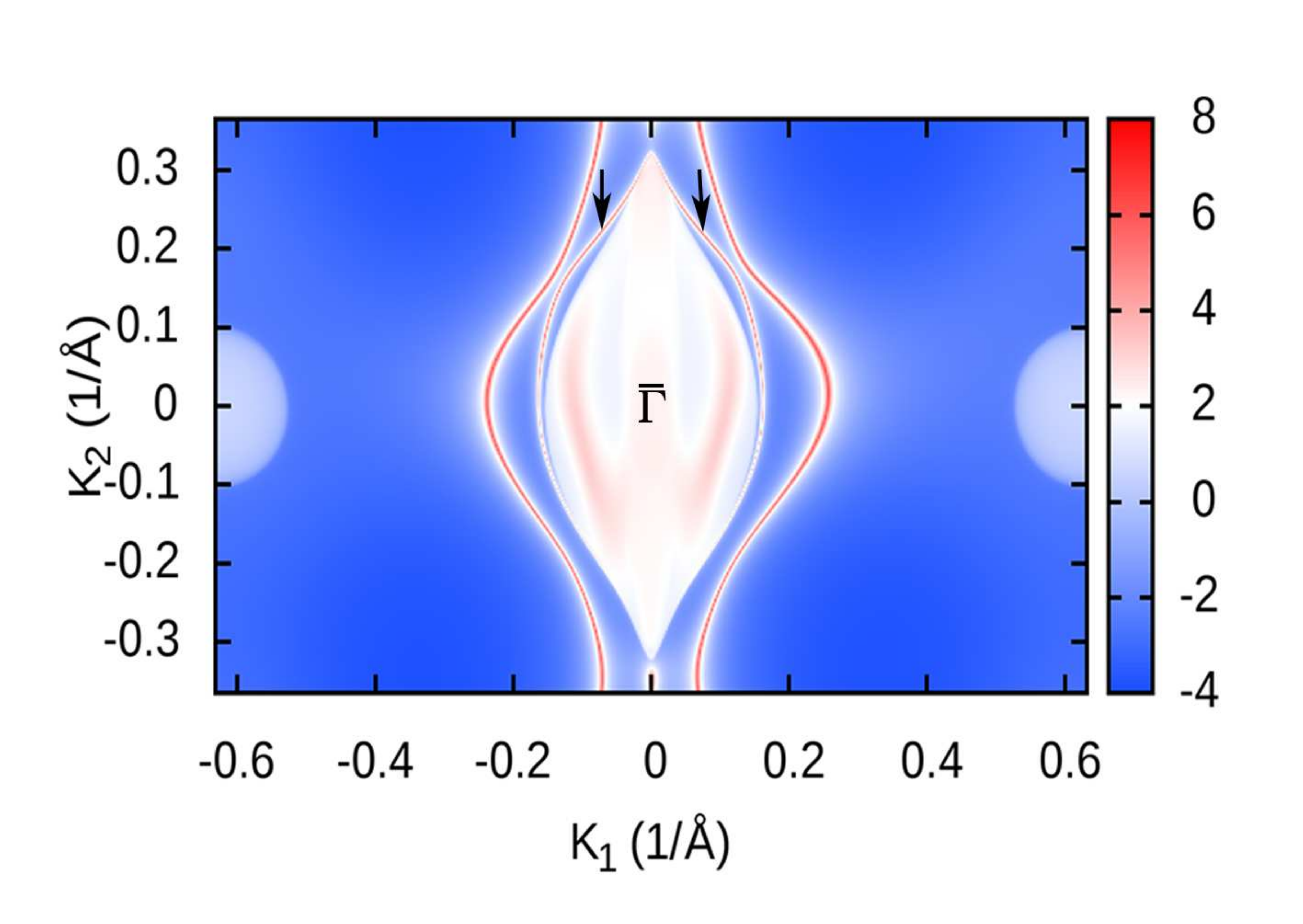}
	\includegraphics[width=.49\textwidth]{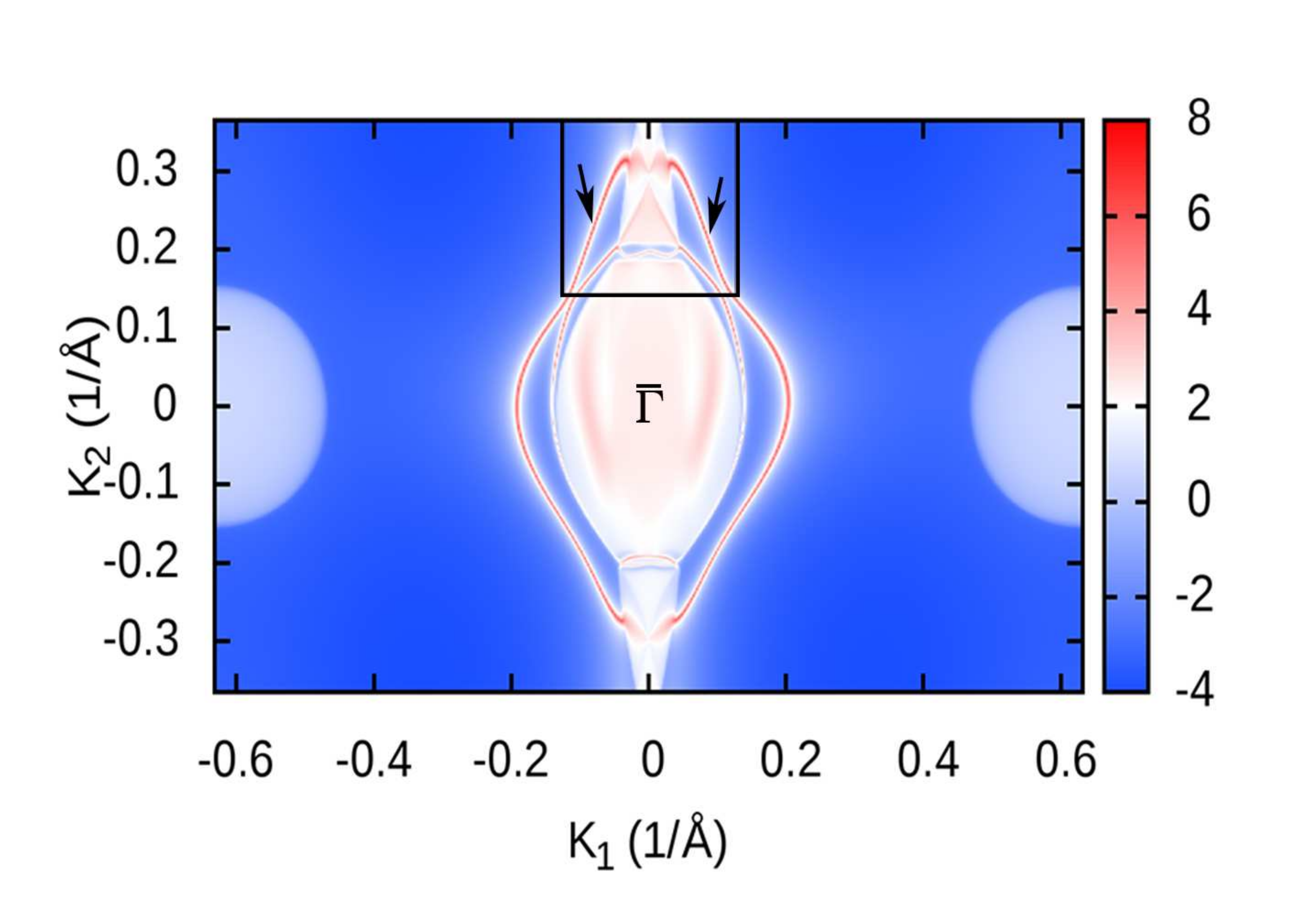}
	\includegraphics[width=.49\textwidth]{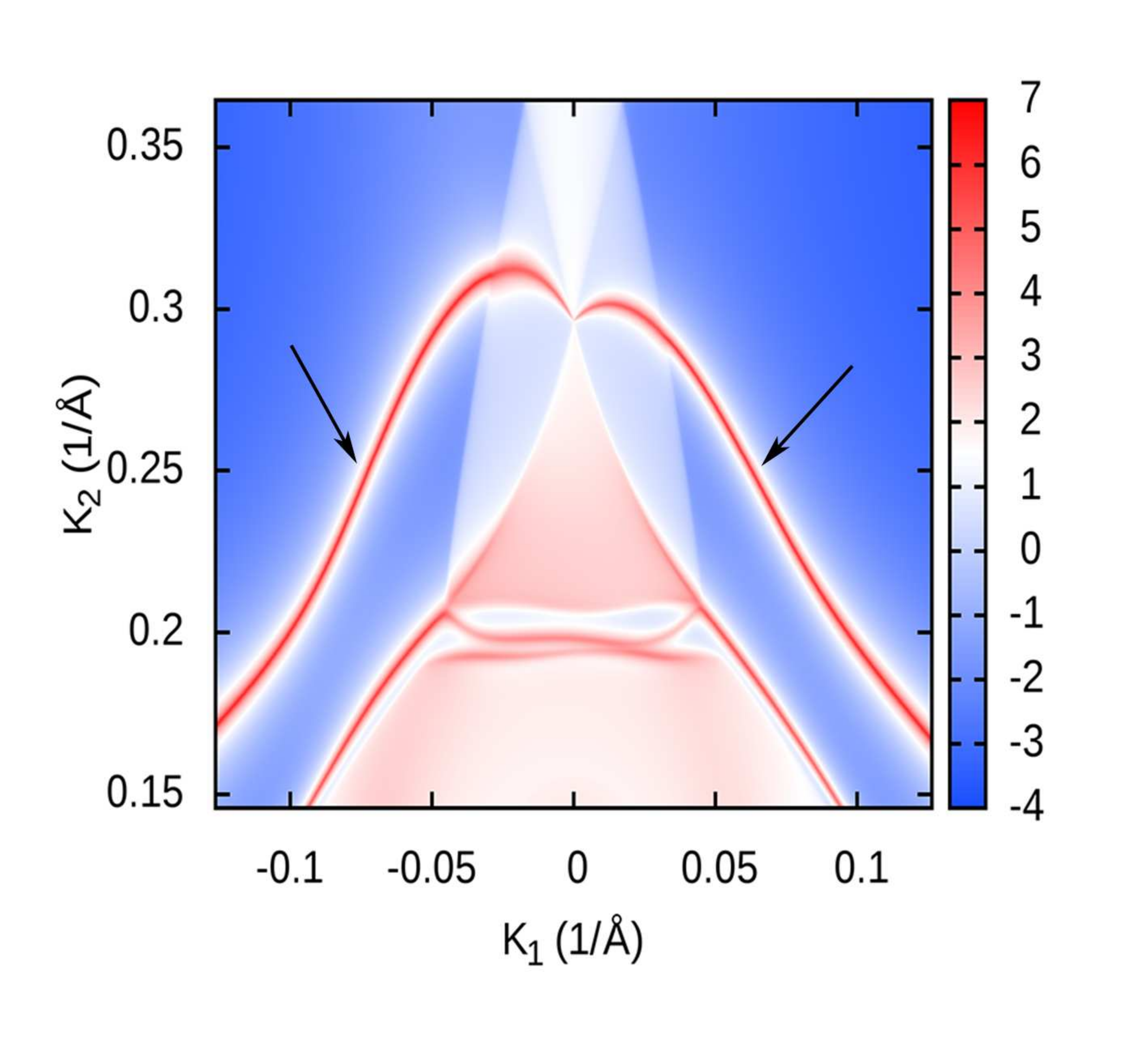}
	\includegraphics[width=.49\textwidth]{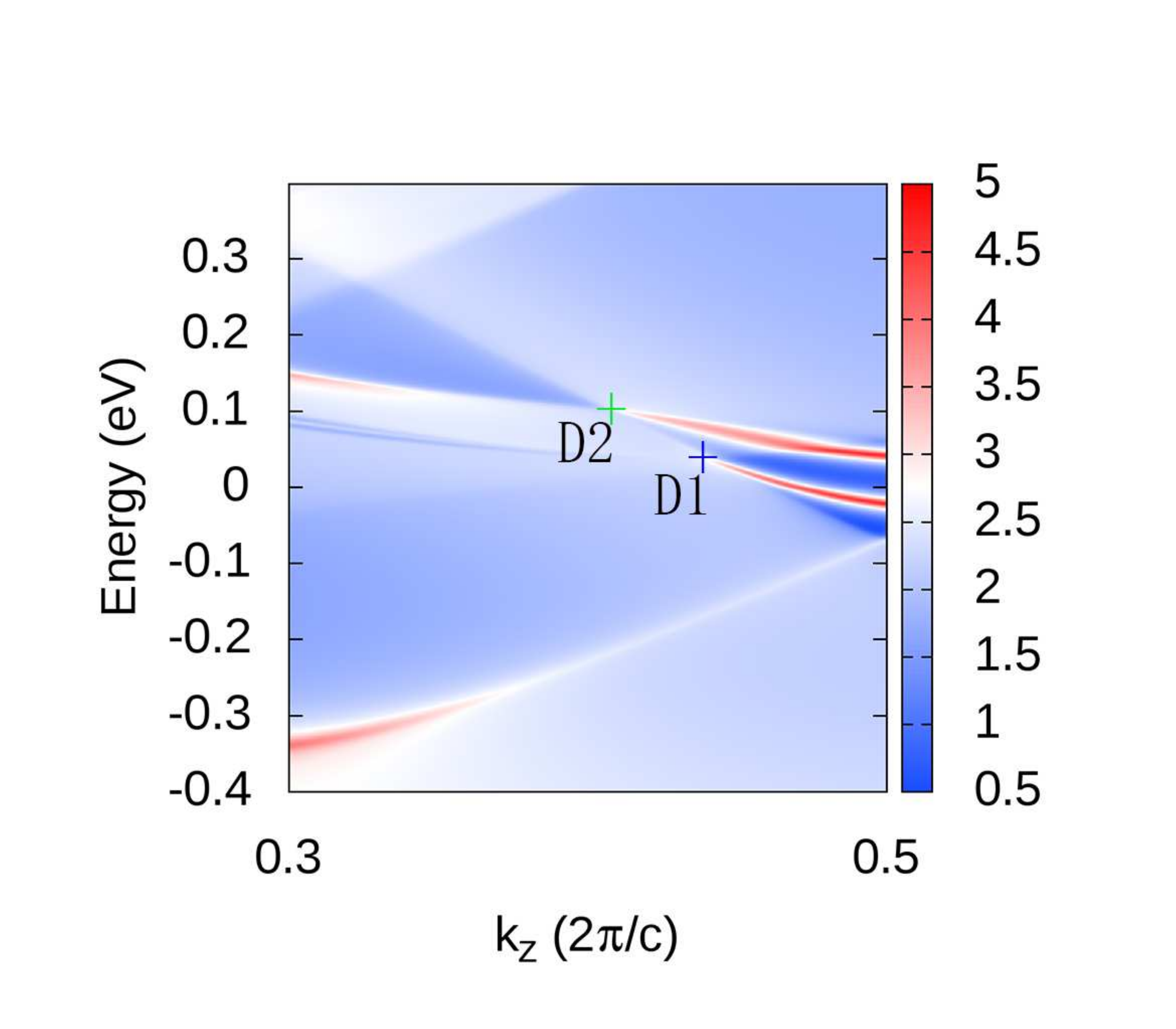}
	\caption{Surface Fermi surface with SOC of (a) at the energy of type-\uppercase\expandafter{\romannumeral4} DPs and (b) at the energy of type-\uppercase\expandafter{\romannumeral2} DPs. (c) An enlarged view of the area highlighted by the black box in (b). (d) Surface band structure of SrAgBi along the $\bar{\Gamma}$-$\bar A$ direction on the $(100)$ surface BZ.}
	\label{surface state}
\end{figure}

Fermi arc electron states on the surface of the crystal are the signature of the Dirac semimetal state. In our case, we present calculations of the (100) surface states. Figures \ref{surface state}(a) and \ref{surface state}(b) are (100) Fermi surfaces for type-\uppercase\expandafter{\romannumeral4} and type-\uppercase\expandafter{\romannumeral2} DPs. There exist the Fermi arcs marked by the arrows and the Fermi arcs are terminating at the projected DPs. Figure \ref{surface state}(c) shows the enlarged view of the area highlighted by the black box in Fig. \ref{surface state}(b). Figure \ref{surface state}(d) shows the energy dispersion of the surface band structure along $\bar{\Gamma}$-$\bar A$ ($k_z$) direction. The projected Dirac points are denoted as a blue ``+" and a green ``+" respectively.  We observe surface states that emerge out of the DPs at $k_z$=0.406($2\pi/c$) and $k_z$=0.438($2\pi/c$).

\section{MODELS}

\subsection{Effective model of type-\uppercase\expandafter{\romannumeral4} Dirac fermions}
\begin{figure}[h]
	\centering
	\subfigure[]{
		\begin{minipage}[t]{0.46\linewidth}
			\centering
			\includegraphics[width=3.5cm]{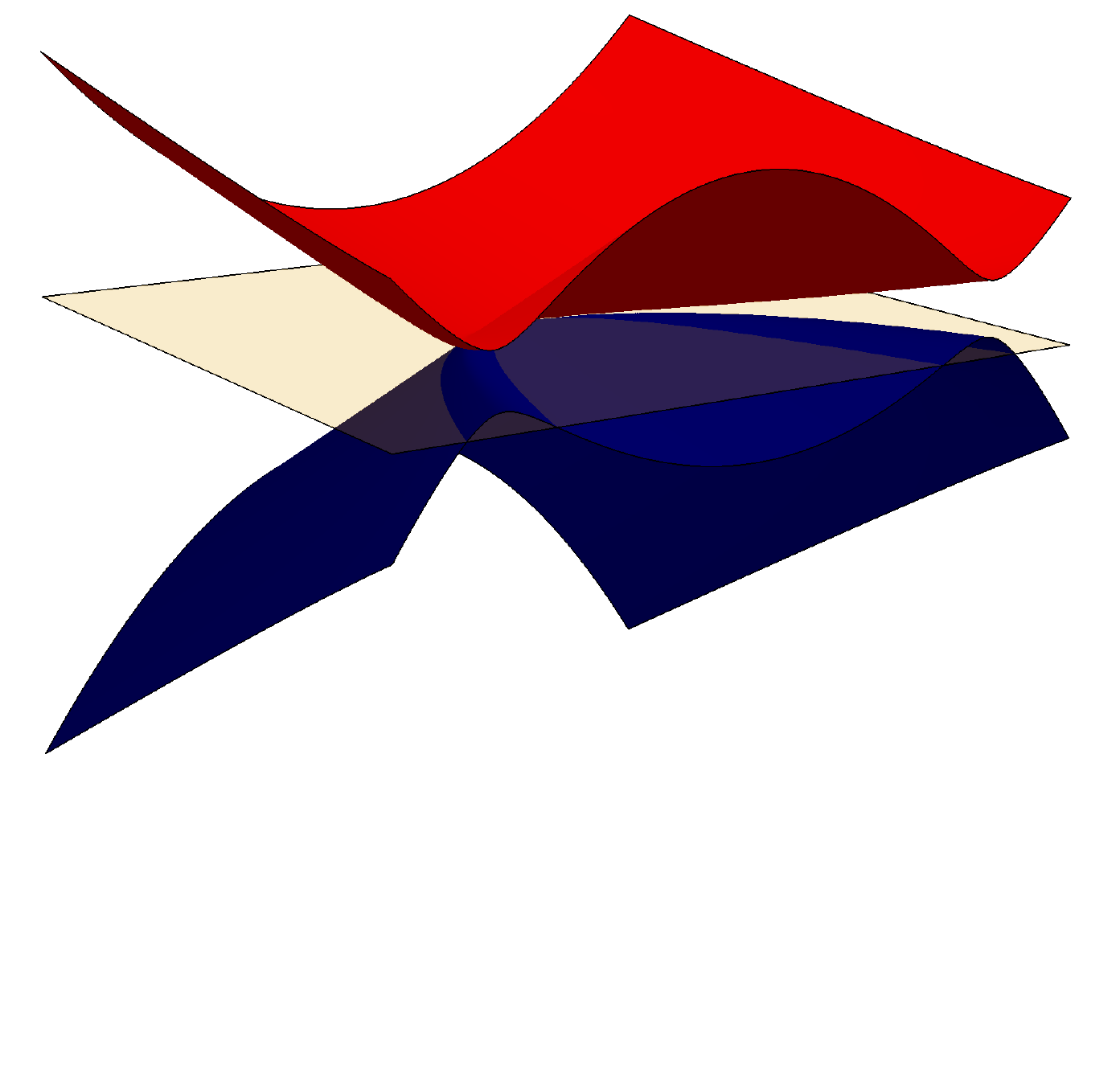}
		\end{minipage}
	}
	\subfigure[]{
		\begin{minipage}[t]{0.46\linewidth}
			\centering
			\includegraphics[width=3.5cm]{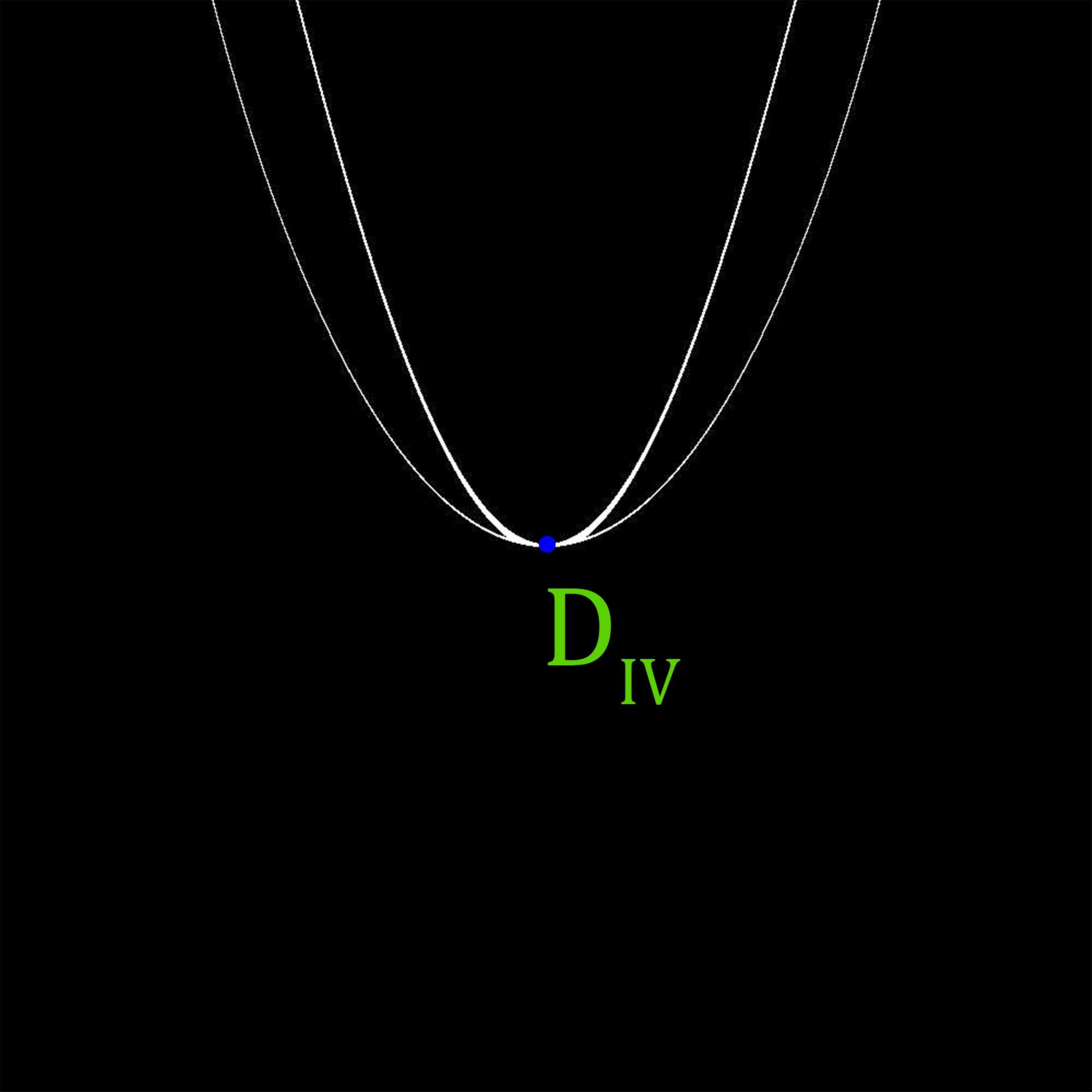}
		\end{minipage}
	}
	\centering
	\caption{(a)  Dispersion around the type-\uppercase\expandafter{\romannumeral4} DP by $k{\cdot}p$ model fitting. (b) Fermi surface of type-\uppercase\expandafter{\romannumeral4} DSM.}
	\label{type 4}
\end{figure}

To characterize the type-\uppercase\expandafter{\romannumeral4} Dirac fermion (marked by D1 in Fig. \ref{bands and pocket}(c)), we construct a $k{\cdot}p$ effective model around it, subjected to the symmetry constraints~\cite{wang2012dirac,bradley2009mathematical,mardanya2019prediction}. On $\Gamma$-A path, the point group symmetry is of $C_{6v}$. Using the $\Delta_{9}$ and $\Delta_{7}$ states of the $C_{6v}$ group as the basis with components $\Delta_{9}(+3/2)$, $\Delta_{9}(-3/2)$, $\Delta_{7}(+1/2)$, $\Delta_{7}(-1/2)$, the Hamiltonian around D1 to $\mathcal{O}(k_{x,y}^2)$ and $\mathcal{O}(k_z)$ is given by,

\begin{equation}
	H(\mathbf{k}) =
	\left[
	\begin{array}{cccc}
		\epsilon_{1}(\mathbf{k}) & 0 & Ck_{+}^{2} & D(k_z)k_{+} \\
		0 & \epsilon_{1}(\mathbf{k}) & -D(k_z)k_{-} & Ck_{-}^{2} \\
		Ck_{-}^{2} & -D(k_z)k_{+} & \epsilon_{2}(\mathbf{k}) & 0 \\
		D(k_z)k_{-} & Ck_{+}^{2} & 0 & \epsilon_{2}(\mathbf{k})
	\end{array}
	\right ],
\end{equation}
where $k_{\pm}=k_{x}\ {\pm}\ ik_{y}$, $\epsilon_{1}(\mathbf{k})=A_{1}k_{+}k_{-}+B_{1}k_{z}$, $\epsilon_{2}(\mathbf{k})=A_{2}k_{+}k_{-}+B_{2}k_{z}$, and $D(k_{z})=D_{0}+D_{1}k_{z}$. The eigenvalues of the above Hamiltonian are,
\begin{equation}
	\begin{aligned}
		&E(\mathbf{k})  = \frac{\epsilon_{1}(\mathbf{k})+\epsilon_{2}(\mathbf{k})}{2} \\
		&\ \ \ \ \pm\sqrt{\left(\frac{\epsilon_{1}(\mathbf{k})-\epsilon_{2}(\mathbf{k})}{2}\right)^2+C^{2}(k_{+}k_{-})^{2}+D(k_{z})^{2}k_{+}k_{-}}.
	\end{aligned}
\end{equation}
By fitting the energy spectrum of the effective Hamiltonian with that of the $ab\ initio$ calculation, the parameters in the effective model can be determined. For type-\uppercase\expandafter{\romannumeral4} Dirac fermions in SrAgBi, our fitting leads to $A_{1}=-344.688\ \mathrm{eV\AA}^{2}$, $B_{1}=-1.45388\ \mathrm{eV\AA}$, $A_{2}=238.175\ \mathrm{eV\AA}^2$, $B_{2}=0.437982\ \mathrm{eV\AA}$, $C=-0.009549\ \mathrm{eV\AA}^2$, $D_{0}=0.000635\ \mathrm{eV\AA}$, $D_{1}=44.7332\ \mathrm{eV\AA}^2$. The band at the energy of type-\uppercase\expandafter{\romannumeral4} DP in $k_y{-}k_z$ plane and the corresponding Fermi surface are shown in Fig. \ref{type 4}. One can find that the Fermi surface is like two quadratic functions touching at the type-\uppercase\expandafter{\romannumeral4} Dirac point.

\subsection{Coexistence of type-\uppercase\expandafter{\romannumeral2} and type-\uppercase\expandafter{\romannumeral4} Dirac fermions}

Now we begin to illustrate the coexistence of  type-\uppercase\expandafter{\romannumeral2} and type-\uppercase\expandafter{\romannumeral4} DSM phase in a simplified model.  Consider two pairs of DPs located on $\Gamma$-A in the BZ. Here, $\Gamma$-A is the principal rotation axis, which offers the symmetry protection needed for the DPs~\cite{yang2014classification}. Because the nontrivial band inversion of SrAgBi is determined by the low-energy bands along $\Gamma$-A, while the bands elsewhere are far away from Fermi level. To describe SrAgBi, we need an eight-band model, because of the following considerations. (\romannumeral1) There are two pairs of bands near the Fermi level. (\romannumeral2) One band crosses the other two bands with different symmetry characters to form the DPs. (\romannumeral3) Each band should be Kramers degenerate in the presence of both time-reversal($\mathcal T$) and inversion($\mathcal P$).

With the reasons above, we consider the following model defined around $\Gamma$-A~\cite{zhu2019composite}:
\begin{equation}\label{eq:eight band kp model}
	\mathcal{H}_{\mathrm{eff}} =
	\left[
	\begin{array}{cc}
		H_{\uparrow \uparrow} & \mathbf{0} \\
		\mathbf{0} & H_{\downarrow \downarrow}
	\end{array}
	\right ],
\end{equation}
where $H_{\uparrow \uparrow}$ and $H_{\downarrow \downarrow}$ are $4 \times 4$  matrices with
\begin{equation}
	H_{\uparrow \uparrow} =
	\left[
	\begin{array}{cccc}
		M_{1} & B_{1}\mathrm{cos}\frac{k_z}{2} & 0 & Ak_{+} \\
		B_{1}\mathrm{cos}\frac{k_z}{2} & M_1 & Ak_{+} & 0 \\
		0 & Ak_{-} & M_2 & B_{2}\mathrm{cos}\frac{k_z}{2} \\
		Ak_{-} & 0 & B_{2}\mathrm{cos}\frac{k_z}{2} & M_2
	\end{array}
	\right ]
\end{equation}
and $H_{\downarrow \downarrow} = H_{\uparrow \uparrow}^{\ast}$. Here, $k_{\pm} = k_x \pm ik_y$ and $A$, $B_i$, and $M_i$ $(i = 1, 2)$ are the parameters of the model. On the $\Gamma$-A path in which $k_{\pm} = 0$, one obtains the following Kramers degenerate spectrum:
\begin{equation}
	\varepsilon_{i,\pm}(k_z) = M_i \pm B_i\mathrm{cos}\frac{k_z}{2}.
\end{equation}
At $\Gamma$ and A,  the band energies respectively are
\begin{equation}
	\varepsilon_{i,\pm}^{\Gamma} = M_i \pm B_i, \qquad\varepsilon_{i,\pm}^{A} = M_i.
\end{equation}

\begin{figure}[htp]
	\centering
	\subfigure[]{
		\begin{minipage}[t]{0.46\linewidth}
			\centering
			\includegraphics[width=4cm]{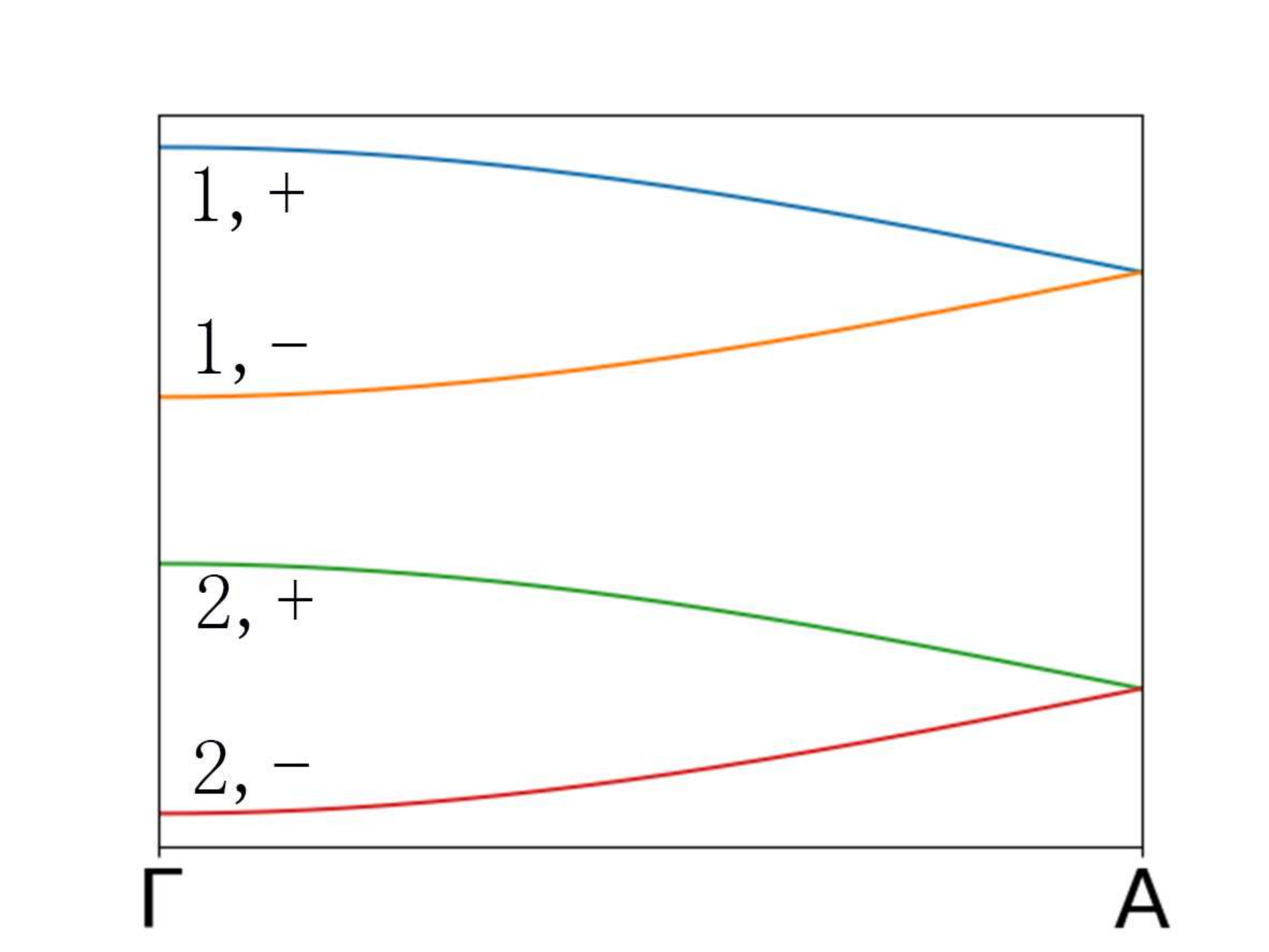}
		\end{minipage}
	}
	\subfigure[]{
		\begin{minipage}[t]{0.46\linewidth}
			\centering
			\includegraphics[width=4cm]{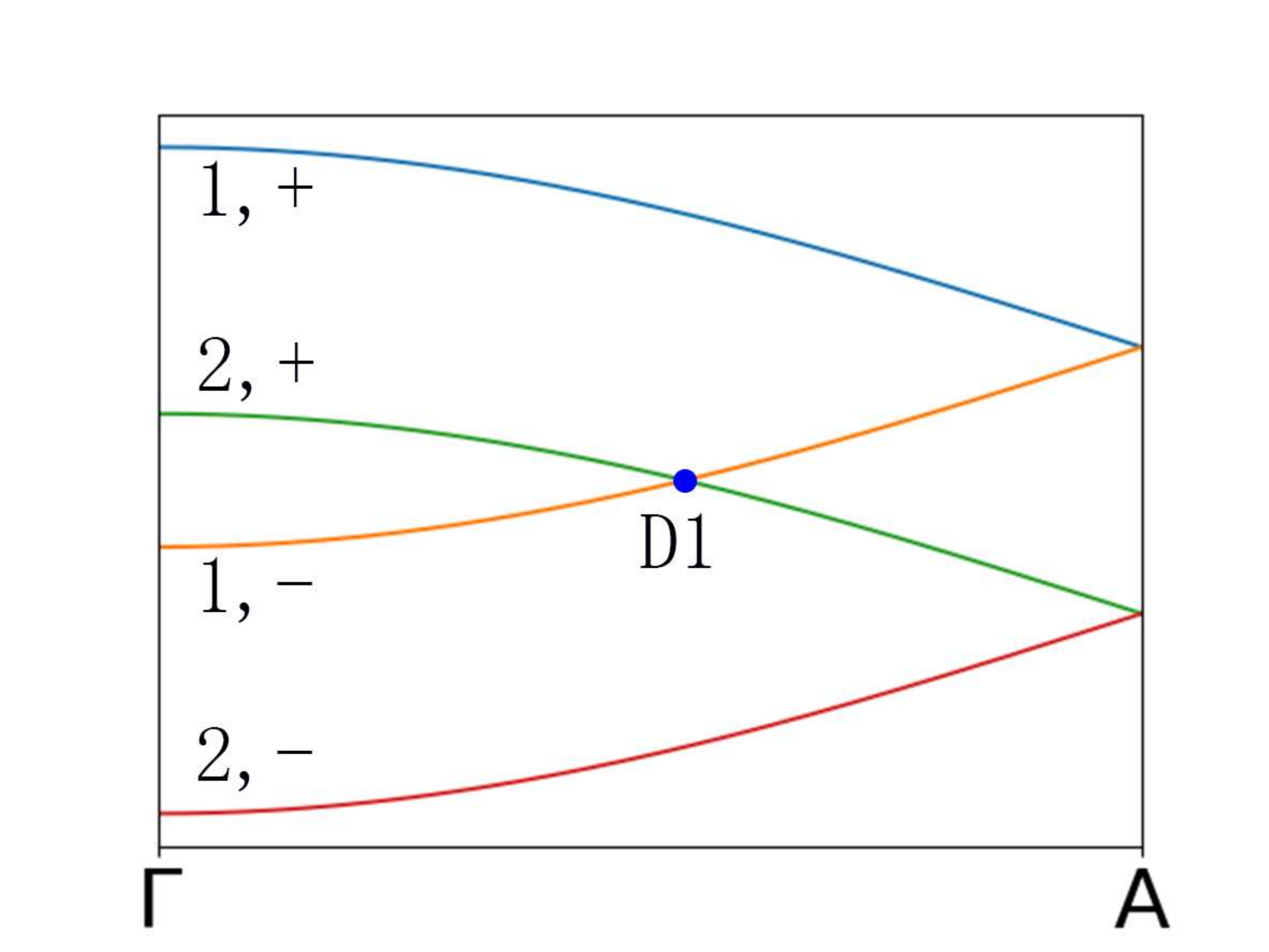}
		\end{minipage}
	}
	\subfigure[]{
		\begin{minipage}[t]{0.8\linewidth}
			\centering
			\includegraphics[width=6cm]{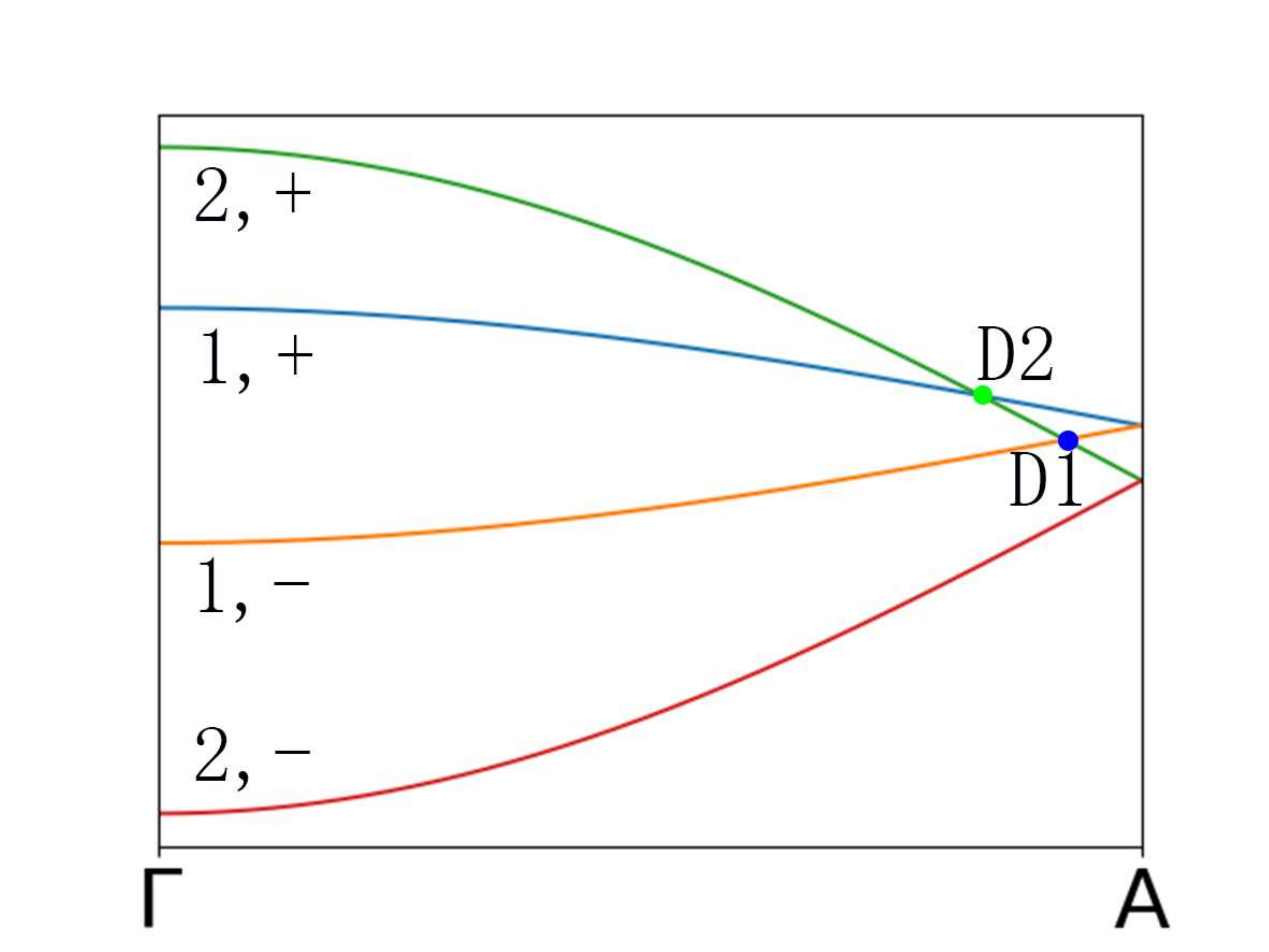}
		\end{minipage}
	}
	\centering
	\caption{Three types of phases with distinct band ordering along the $\Gamma$-$A$ path, as described by the Eq. (\ref{eq:eight band kp model}). (c) is consistent with the band of SrAgBi along $\Gamma$-$A$ near the Fermi energy.}
	\label{eight band kp pic}
\end{figure}

By tuning $M_i$ and $B_i$, the band structures for equation (\ref{eq:eight band kp model}) can drop into three distinct cases, as shown in Fig. \ref{eight band kp pic}. We assume that in the atomic limit the pair $\varepsilon_{1,\pm}$ is energetically above $\varepsilon_{2,\pm}$, and consider the bands are half-filled. The case in Fig. \ref{eight band kp pic}(a) is a trivial insulating phase adiabatically connected to the atomic limit. For the case in Fig. \ref{eight band kp pic}(b), there is only one band inversion at $\Gamma$, leading to the formation of DPs. For the case in Fig. \ref{eight band kp pic}(c), there are two band inversions at $\Gamma$, this situation is consistent with SrAgBi.

\begin{figure}[htp]
	\centering
	\subfigure[]{
		\begin{minipage}[t]{0.3\linewidth}
			\centering
			\includegraphics[width=3.0cm]{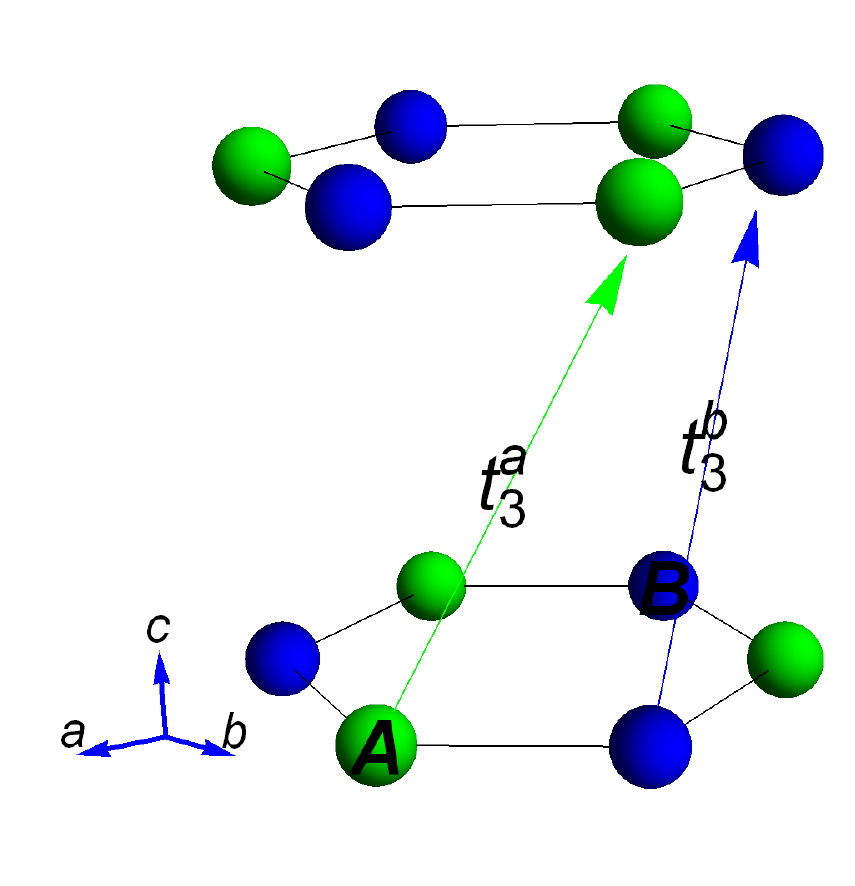}
		\end{minipage}
	}
	\subfigure[]{
		\begin{minipage}[t]{0.4\linewidth}
			\centering
			\includegraphics[width=4.0cm]{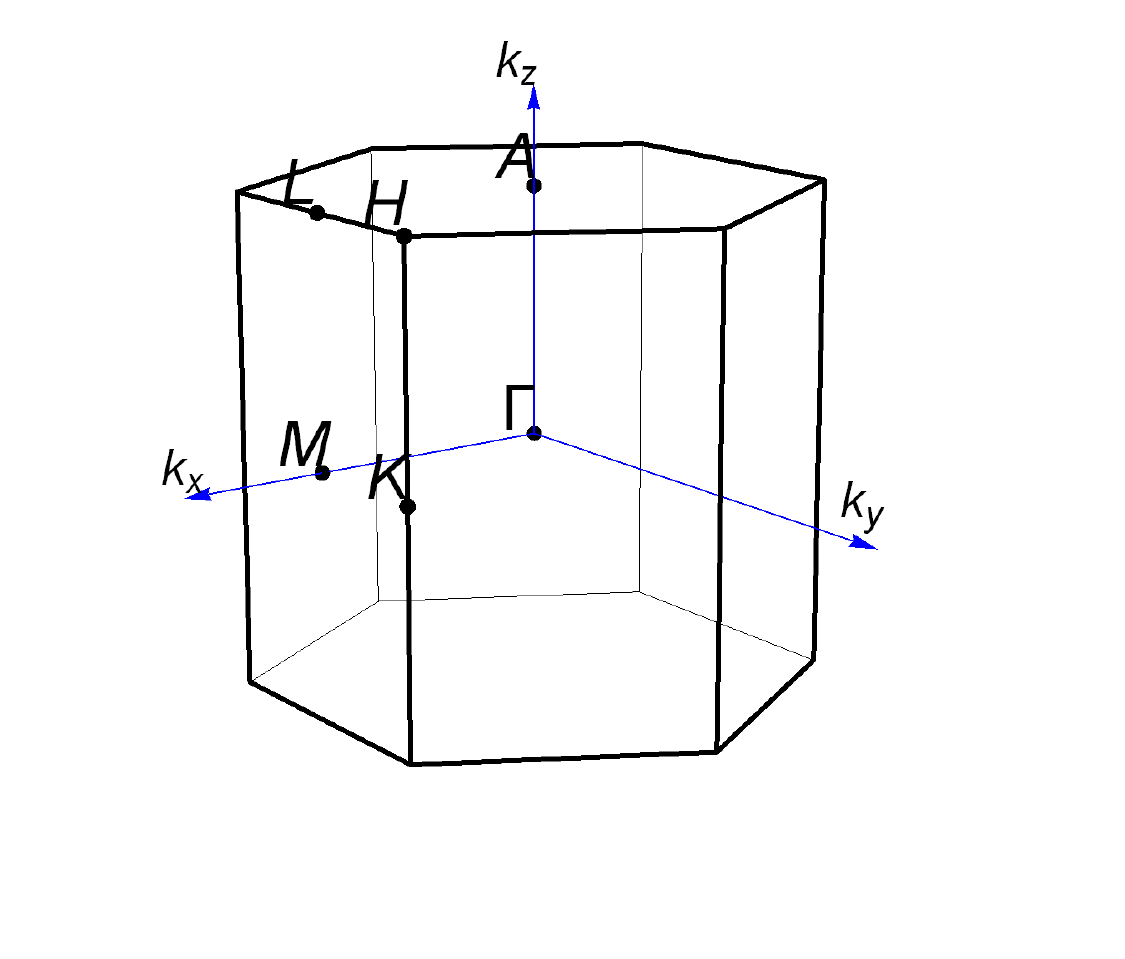}
		\end{minipage}
	}
	\subfigure[]{
		\begin{minipage}[t]{0.9\linewidth}
			\centering
			\includegraphics[width=6cm]{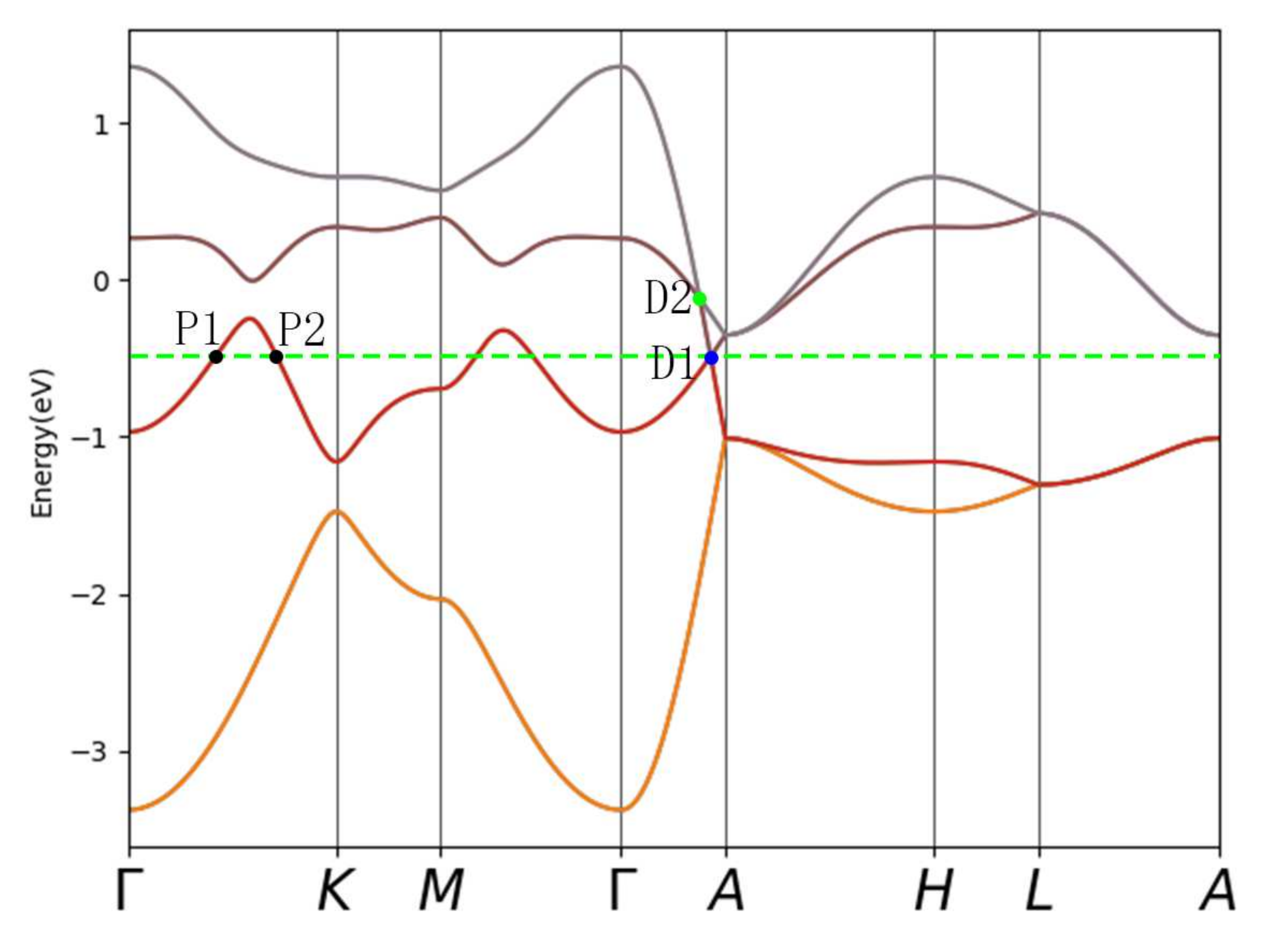}
		\end{minipage}
	}
	\subfigure[]{
		\begin{minipage}[t]{0.9\linewidth}
			\centering
			\includegraphics[width=6.0cm]{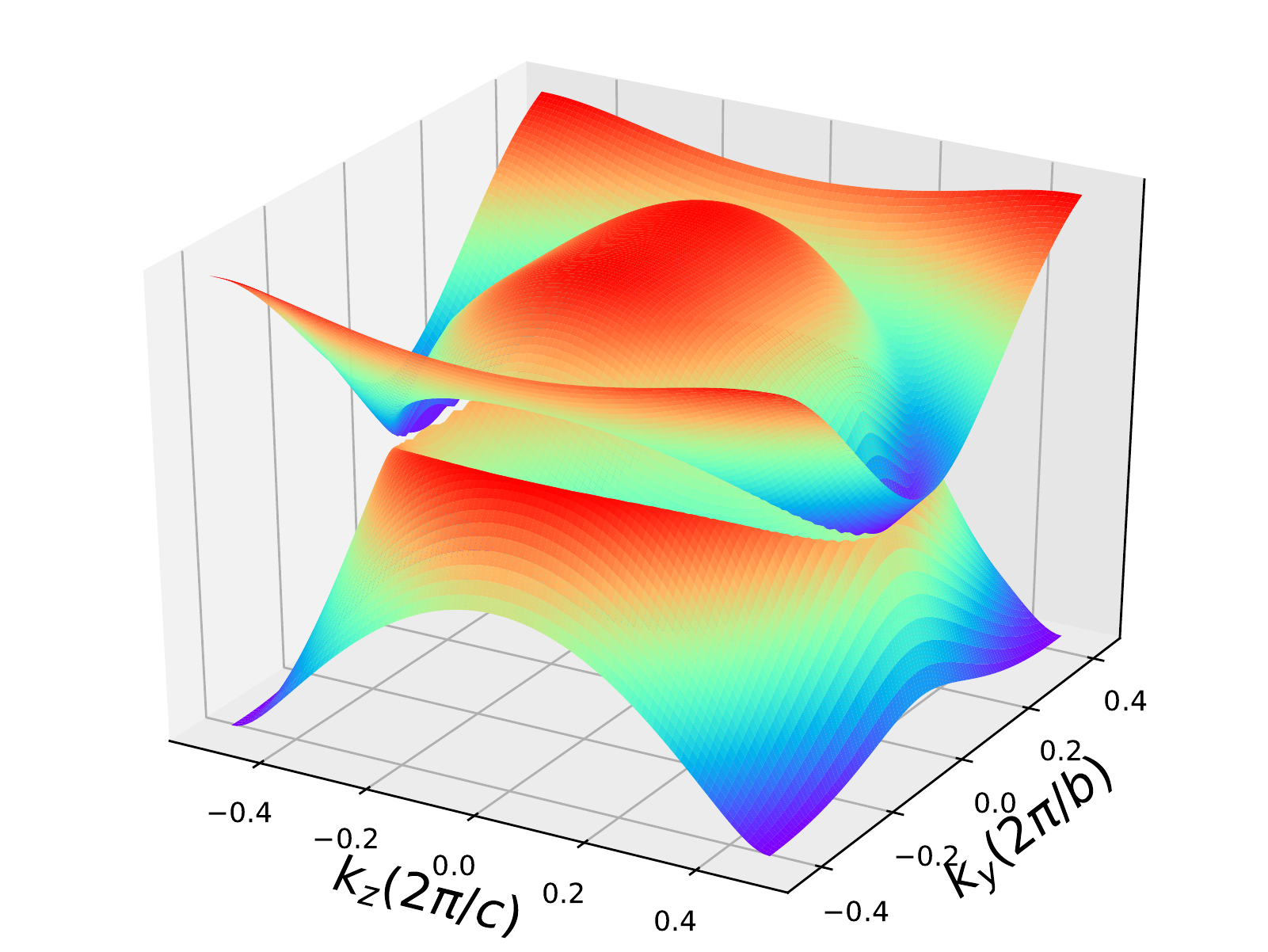}
		\end{minipage}
	}
	\centering
	\caption{(a) 3D lattice model. The arrows indicate two interlayer hopping processes. (b) The bulk BZ of this model. (c) Band structure of the model. (d) 3D band structure on ${k_y}-{k_z}$ plane; we have used $\epsilon_a = -1.1083$ eV, $\epsilon_b = 0.1087$ eV, $t_1 = 0.2533$ eV, $t_2^a=0.0166$ eV, $t_2^b = -0.0766$ eV, $t_3^a = -0.3950$ eV, $t_3^b = -0.1030$ eV.}
	\label{tbmodel pic}
\end{figure}

The above analysis provides an intuitive picture of the band inversion and band crossing along $\Gamma$-A of SrAgBi, yet the symmetry protections cannot be captured within the simplified model. To fully characterize the type-\uppercase\expandafter{\romannumeral4} DSM phase, we extend the simplified equation (\ref{eq:eight band kp model}) to a tight-binding model. SrAgBi has three important symmetries in addition to the $\mathcal{T}$ and $\mathcal{P}$: A sixfold screw rotation ${\tilde C_6} : (x,y,z) \rightarrow (x/2-\sqrt{3}y/2,\sqrt{3}x/2+y/2,z+1/2)$, a horizontal mirror $M_z : (x,y,z) \rightarrow (x,y,-z+1/2)$, and a vertical glide mirror ${\tilde M_y} : (x,y,z) \rightarrow (x,-y,z+1/2)$. According to the first principle calculation, we find that almost all of the electron state near the fermi energy come from $Ag$'s $d$ orbital and $Bi$'s $p$ orbital. So we consider a 3D lattice consisting of 2D honeycomb layers stacked along $z$, as sketched in Fig. \ref{tbmodel pic}(a). For each layer, the $A$ and $B$ sites are occupied by two different types of atoms, $Ag$ and $Bi$, respectively. Each unit cell contains two layers, between which $A$ and $B$ are switched. We assume that each site has two basis orbitals forming a Kramers pair: $\vert p_{+},\uparrow\rangle$ and $\vert p_{-},\downarrow\rangle$ on $A$, whereas $\vert d_{+2},\uparrow\rangle$ and $\vert d_{-2},\downarrow\rangle$ on $B$. where $p_{\pm} = p_x \pm ip_y$ and $d_{\pm 2} = d_{x^2-y^2} \pm 2id_{xy}$. Based on these, we use the following tight-binding model~\cite{zhu2019composite}:
\begin{equation}\label{eq:eightband tight binding model}
	\begin{aligned}
		\mathcal{H}&=\sum_{\alpha,i}(\epsilon_{a}a_{\alpha,i}^{\dag}a_{\alpha,i}+\epsilon_{b}b_{\alpha,i}^{\dag}b_{\alpha,i})\\
		&\quad +\sum_{\alpha,i,m}t_1(-1)^{\alpha}(a_{\alpha,i+\mathbf{R}_m}^{\dag}\sigma_{z}e^{i(2m-1)\pi/3}b_{\alpha,i}+\mathbf{H.c.}) \\
		&\quad +\sum_{\alpha,i,n}(t_{2}^{a}a_{\alpha,i+\mathbf{R}_n^{'}}^{\dag}a_{\alpha,i}+t_{2}^{b}b_{\alpha,i+\mathbf{R}_n^{'}}^{\dag}b_{\alpha,i})\\
		&\quad +\sum_{i,m}(t_{3}^{a}a_{0,i+\mathbf{R}_m}^{\dag}a_{1,i}+t_{3}^{b}b_{1,i+\mathbf{R}_m}^{\dag}b_{0,i}+\mathbf{H.c.}).
	\end{aligned}
\end{equation}

\begin{figure}[htp]
	\centering
	\subfigure[]{
		\begin{minipage}[t]{0.9\linewidth}
			\centering
			\includegraphics[width=4.5cm,height=4cm]{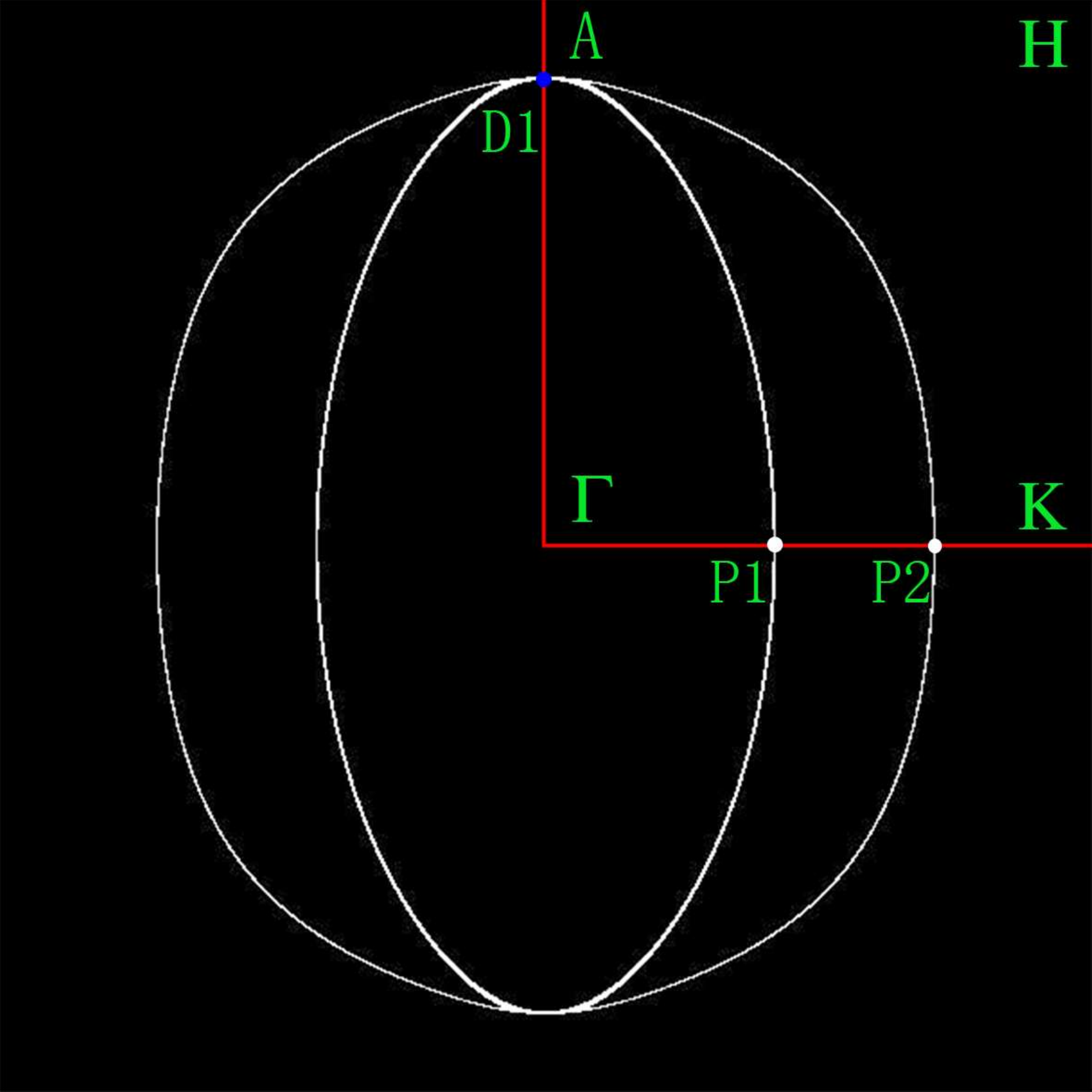}
		\end{minipage}
	}
	\subfigure[]{
		\begin{minipage}[t]{0.9\linewidth}
			\centering
			\includegraphics[width=4.5cm]{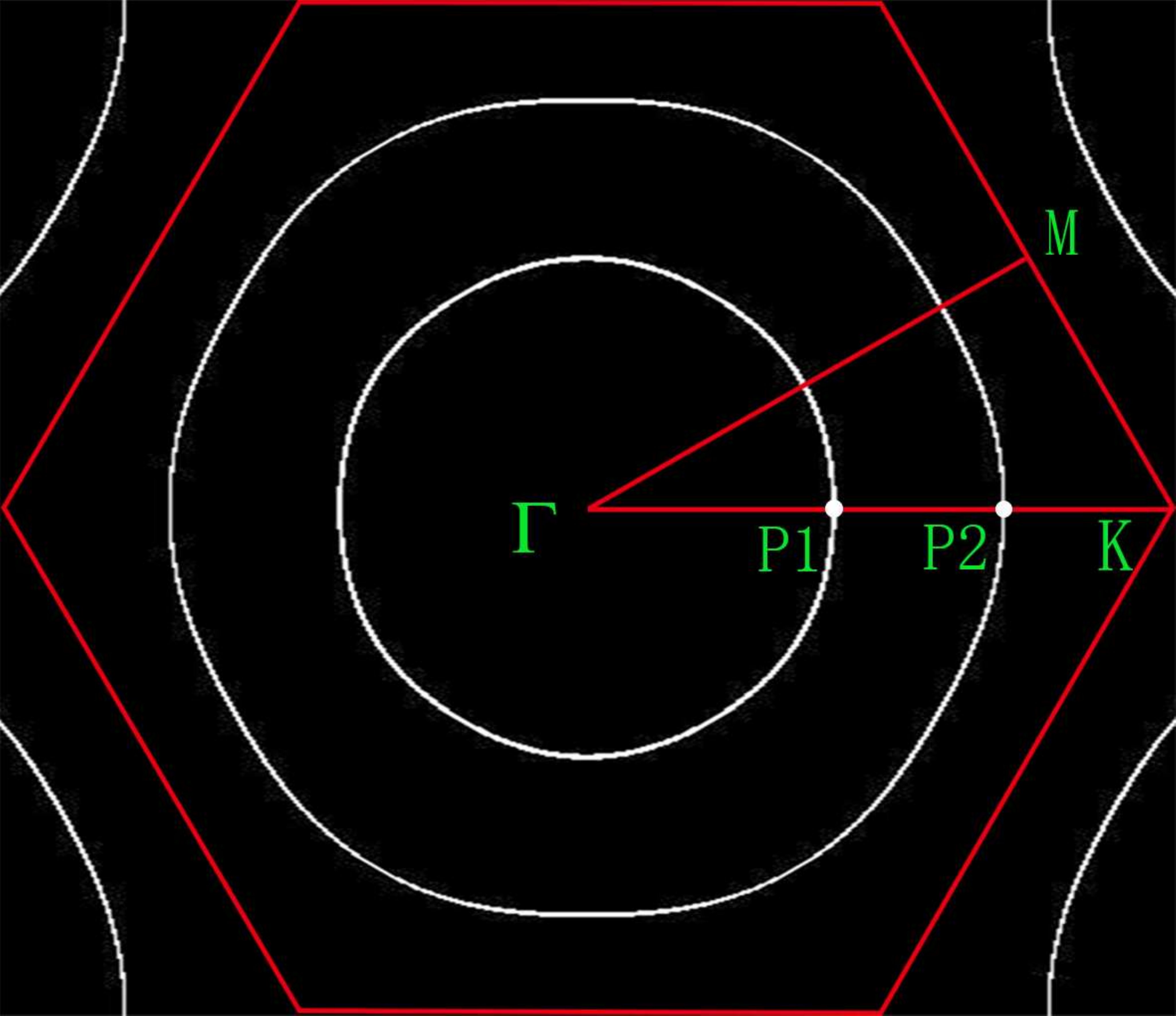}
		\end{minipage}
	}
	\centering
	\caption{(a) ${k_y}-{k_z}$ and (b) ${k_x}-{k_y}$ plane bulk-state equal-energy contours.}
	\label{tbmodel pocket}
\end{figure}

Here, $a^{\dag}=(a_{\vert p_{+},\uparrow\rangle}^{\dag},a_{\vert p_{-},\downarrow\rangle}^{\dag})$ and $b^{\dag}=(b_{\vert d_{+2},\uparrow\rangle}^{\dag},b_{\vert d_{-2},\downarrow\rangle}^{\dag})$ are the electron creation operators, $\alpha = 0, 1$ label the two layers in a unit cell, $i$ labels the sites within a layer, $\mathbf{R}_m$ $(m = 1, 2, 3)$ correspond to the vectors connecting to the three nearest neighbors in a layer, $\mathbf{R}_n^{'}$ $(n = 1,...,6)$ correspond to the vectors connecting to the six next-nearest neighbors in a layer, $\epsilon_a$ and $\epsilon_b$ are the on-site energies, and the $t$'s are various hopping amplitudes. In model (\ref{eq:eightband tight binding model}), the first term represents an on-site energy difference, and the second and third terms are hoppings within a honeycomb layer. The last term represents the strongest interlayer hopping. This model retains the main symmetry of SrAgBi.  If the Sr atoms in SrAgBi were removed, the crystal lattice would become identical to the lattice for this tight-binding model.

Figure \ref{tbmodel pic}(c) shows the coexistence of the type-\uppercase\expandafter{\romannumeral2} and type-\uppercase\expandafter{\romannumeral4} DSM phase in the tight-binding model (\ref{eq:eightband tight binding model}), and the low-energy physics resembles that in Fig. \ref{bands and pocket}(b). One can see that there is part of valance band (P1-P2) along $\Gamma$-K in Fig. \ref{tbmodel pic}(c), which is higher than the Fermi level, this is very important in forming type-\uppercase\expandafter{\romannumeral4} DPs. The band near the type-\uppercase\expandafter{\romannumeral4} DP (D1) is shown in Fig. \ref{tbmodel pic}(d), which is consistent with Fig. \ref{bands and pocket}(d).  Figures \ref{tbmodel pocket}(a) and \ref{tbmodel pocket}(b) show the bulk-state equal-energy contours of ${k_y}-{k_z}$ and $k_z = 0$ plane, this means that there is a electron pocket in a hole pocket and these two pockets touch at type-\uppercase\expandafter{\romannumeral4} DPs. One can find Fig. \ref{tbmodel pocket}(a) is similar to Fig. \ref{2d pocket 6}(b), this proves that the model we chose is effective in describing SrAgBi.

\section{Conclusion}
By using the first principle calculation with SOC, we proposed that type-\uppercase\expandafter{\romannumeral4} Dirac fermions exist in the bulk state of SrAgBi. Distinct from type-\uppercase\expandafter{\romannumeral1}, type-\uppercase\expandafter{\romannumeral2} and type-\uppercase\expandafter{\romannumeral3} Dirac semimetals such as Na$_3$Bi~\cite{wang2012dirac}, VGa$_3$~\cite{chang2017type} and Zn$_2$In$_2$S$_5$~\cite{huang2017black}, SrAgBi has nonlinear bands near the type-\uppercase\expandafter{\romannumeral4} DPs. The $k{\cdot}p$ model shows that the type-\uppercase\expandafter{\romannumeral4} Dirac fermion is a kind of higher-order Dirac fermion~\cite{wu2020higher}. Because there is a type-\uppercase\expandafter{\romannumeral2} DP near the type-\uppercase\expandafter{\romannumeral4} DP, SrAgBi provides a platform for exploring the interplay between type-\uppercase\expandafter{\romannumeral2} and type-\uppercase\expandafter{\romannumeral4} Dirac fermions, we used two models to describe the coexistence of these two Dirac fermions. Topological surface states of these two DPs are also calculated. The bulk DPs and the surface states can be experimentally probed by the angle-resolved photoemission spectroscopy (ARPES)~\cite{liu2014discovery,xu2015discovery}. In addition, some articles point out that by doping other atoms, one DP in this type of material can be split into two triple points~\cite{mondal2018broken,mardanya2019prediction}, which points out the direction for our future work.

\section*{References}

 \bibliographystyle{unsrt}
 \bibliography{ref}
\end{document}